\documentclass[twocolumn,twocolappendix]{aastex63}
\usepackage{csvsimple}
\usepackage{amsmath}
\usepackage{booktabs}
\usepackage[flushleft]{threeparttable}
\usepackage{hyperref}
\usepackage{xcolor}
\usepackage{comment}
\usepackage[inline]{enumitem}
\setcitestyle{authoryear,open={(},close={)}}

\graphicspath{{./}{Figures/}}

\def\HI{\rm H\,\textsc{i}}
\def\HIspace{\rm H\,\textsc{i} }

\begin{document}

\title{When and how does ram pressure stripping in low-mass satellite galaxies enhance star formation}

\correspondingauthor{Jingyao Zhu}
\email{jingyao.zhu@columbia.edu}

\author{Jingyao Zhu}
\affiliation{Department
of Astronomy, Columbia University, New York, NY 10027, USA}

\author{Stephanie Tonnesen}
\affiliation{Center for Computational Astrophysics, Flatiron Institute, New York, NY 10010, USA}

\author{Greg L Bryan}
\affiliation{Department
of Astronomy, Columbia University, New York, NY 10027, USA}
\affiliation{Center for Computational Astrophysics, Flatiron Institute, New York, NY 10010, USA}

\begin{abstract}
We investigate how a satellite’s star formation rate (SFR) and surviving gas respond to ram pressure stripping in various environments. Using a suite of high-resolution “wind-tunnel” simulations with radiative cooling, star formation, and supernovae feedback, we model the first infall orbit of a low-mass disk galaxy ($M_{*} = 10^{9.7}$ $M_{\odot}$) in different host halos, ranging from Milky Way-like to cluster hosts. When the ram pressure is moderate, we find that the stripping satellite shows an enhanced SFR relative to the isolated control case, despite gas loss due to stripping. The SFR enhancement is caused, not directly by compression, but by ram pressure-driven mass flows, which can increase the dense gas fraction in the central disk regions. The spatially-resolved star formation main sequence and Kennicutt-Schmidt relations in our simulations are consistent with recent findings of the VERTICO and GASP surveys. Our results predict the environmental signals of RPS in future multiwavelength, high-angular resolution observations: the star formation and gas surface densities will be centralized, and symmetrically enhanced within the stripping radius.

\end{abstract}

\section{Introduction} \label{sec:intro}

A galaxy can either be star-forming or quenched, depending on internal and environmental processes \citep{kauffmann_environmental_2004,baldry_galaxy_2006,peng_mass_2010}. For the central galaxies within halos, 
internal processes such as supernova and AGN feedback are the main star formation regulators \citep{croton_many_2006,dalla_vecchia_simulating_2008}. This explains the ``main sequence" of star formation over cosmic time: a tight correlation between the star formation rate (SFR) and stellar mass ($M_{*}$) of galaxies \citep{speagle_highly_2014}. For satellite galaxies, environmental factors from the interactions with a central halo become significant: satellites show a strong observational bias to be `red', or star formation quenched, compared with their central counterparts at the same stellar masses \citep{peng_mass_2012,wetzel_galaxy_2012,phillips_dichotomy_2014}. The environmental quenching of satellite galaxies is also ubiquitous in cosmological simulations \citep{tremmel_introducing_2019,wright_quenching_2019,appleby_impact_2020,donnari_quenched_2021,donnari_quenched_2021-1}.

Despite the consensus of environmental quenching in observations and simulations, uncertainties remain in the mass dependence and the scatter of the quenching effectiveness \citep{donnari_quenched_2021}. The uncertainties likely arise from the complex physical processes during the satellite-environment interactions (see recent review by \citealt{cortese_dawes_2021}), of which the dominant is ram pressure stripping (RPS; \citealt{gunn_infall_1972}), the direct removal of the satellite's interstellar medium (ISM) by a host halo medium. RPS galaxies, identified by unidirectional gas tails and little stellar disk deformation, have been observed in several clusters \citep{van_gorkom_interaction_2004,boselli_fate_2006,sun_h_2007,poggianti_jellyfish_2016,deb_gasp_2022} as well as in both idealized (e.g., \citealt{abadi_ram_1999,quilis_gone_2000,schulz_multi_2001,roediger_ram_2006,jachym_gas_2007,mccarthy_ram_2008}) and cosmological simulations \citep{bahe_competition_2012,yun_jellyfish_2019,rohr_jellyfish_2023}. Although the hallmark of RPS is gas removal and, therefore, the eventual quenching of star formation \citep{boselli_fate_2006,crowl_stellar_2008}, recent detailed observations have shown that the early stages of RPS may have complex effects on both the ISM phase distribution and SFRs. Under RPS, the satellite's star formation can be triggered \citep{ebeling_jellyfish_2014,jachym_alma_2019,poggianti_gasp_2019}, the SFR globally \citep{vulcani_enhanced_2018,roberts_lotss_2021,kolcu_quantifying_2022,molnar_westerbork_2022} or locally \citep{vulcani_gasp_2020} enhanced, and the molecular-to-atomic gas ratio boosted \citep{moretti_high_2020}.

The interplay between RPS and star formation is key to understanding environmental quenching, and has been explored in various controlled hydrodynamical simulations. 
Multiple simulations analyzed 
the star formation triggering- or enhancing-potential of RPS 
\citep{schulz_multi_2001,kapferer_effect_2009,tonnesen_star_2012,roediger_star_2014,bekki_galactic_2014,steinhauser_simulations_2016,ruggiero_fate_2017,lee_dual_2020}, but the physical reasons behind the enhancement are unclear. Ram pressure-driven shock passages can trigger local boosts to the SFR, but have little global effects \citep{roediger_star_2014}; pressure enhancement in the wind-leading halves of the galaxies undergoing stripping suggests that compression likely enhances the star formation efficiency \citep{troncoso-iribarren_better_2020}; ram pressure-induced radial gas inflows can modify the star formation morphology, shifting it to the central regions with higher SFR \citep{schulz_multi_2001,tonnesen_star_2012,lee_dual_2020}. There is a need to examine the physical causes of RPS-enhanced star formation in simulations, and to directly compare with recent observations.

In this work, we study the complicated effects of RPS on the ISM distribution and SFRs. (i) We simulate a low-mass spiral galaxy (lowest resolved in \citealt{donnari_quenched_2021}, with tension in the quenched fractions) undergoing RPS in different environments, from a Milky Way-like to a cluster halo, and examine the galaxy's gas and SFR response; (ii) For each host halo, we model a realistic infall orbit with time-varying ram pressure profiles; (iii) We analyze the local SFR-mass relations at comparable spatial resolutions with recent high angular resolution observations \citep{vulcani_gasp_2020,jimenez-donaire_vertico_2023}; and finally (iv) We compare and identify key physical causes of RPS-enhanced star formation.

The structure of this paper is as follows. In \S \ref{sec:methods} we introduce the methodology, with \S \ref{subsec:satellite_galaxy} on the satellite galaxy model, \S \ref{subsec:orbits} on the infall orbits, and \S \ref{subsec:simulations} the simulation initial conditions. We present the simulation global results in \S \ref{sec:global_results}: the time evolution of star formation and the surviving gas (\S \ref{subsec:global_results}), and the gas morphology and kinematics (\S \ref{subsec:gas_kinematics}). Then, \S \ref{sec:resolved_SF_mass} compares the spatially resolved SFR-mass relations ($\Sigma_{\rm SFR} - \Sigma_{\rm gas}$ and $\Sigma_{\rm SFR} - \Sigma_{*}$) between the stripping and isolated galaxy sets. We discuss our results in \S \ref{sec:discussion}: the impact of RPS on star formation (\S \ref{subsec:discussion_RP_SF}), predictions for observations (\S \ref{subsec:obsn_predict}), and limitations of our methodology (\S \ref{subsec:limitations}). \S \ref{sec:conclusion} summarizes the key findings.

\section{Methodology}\label{sec:methods}
We run a suite of three-dimensional ``wind-tunnel" simulations using the adaptive mesh refinement (AMR) code Enzo \citep{bryan_enzo_2014}. The simulation volume is a $162^{3}$ kpc cube with a $128^{3}$ root grid resolution. We allow up to five levels of refinement, giving a highest spatial resolution of $39$ pc (marginally resolving giant molecular clouds). To model the radiative cooling of the multiphase gas, we use the Grackle chemistry and cooling library\footnote{\url{https://grackle.readthedocs.io/}} \citep{smith_grackle_2017}, which calculates photoheating and photoionization from the UV background of \cite{haardt_radiative_2012}. We use the star formation recipe of \cite{goldbaum_mass_2015} with the following parameters: Once a gas cell reaches the Jeans criterion with a number density threshold of $n_{\rm min}=10$ $\rm cm^{-3}$, it forms star particles (including regular stars and Type II supernovae) with a $5\%$ efficiency. The star particles, now followed in our simulations as active particles, subsequently deposit energy into the gas in the forms of stellar and supernovae feedback, under the \cite{goldbaum_mass_2016} feedback model, which includes the terminal momentum input from the number of supernovae expected to go off during a given timestep, adding any additional energy in the form of thermal energy.

We use yt, a multi-code toolkit for analyzing and visualizing astrophysical simulation data \citep{turk_yt_2011}, to create slices and projections, and to select the disk gas and the active star particles for subsequent analyses.

\subsection{The galaxy}\label{subsec:satellite_galaxy}
Our galaxy is placed at the center of the $162$-kpc cubical simulation volume at (81, 81, 81) kpc. We choose a galaxy of low stellar mass $M_{*} = 10^{9.7}$ $M_{\odot}$, motivated by the lowest satellite $M_{*}$ examined in \cite{donnari_quenched_2021} using the IllustrisTNG cosmological simulations \citep{weinberger_simulating_2017,pillepich_simulating_2018}. Table \ref{table:satellite_galaxy} summarizes the global parameters of the satellite galaxy.

\begin{table}
\centering
\caption{Global Parameters of the Low-mass Galaxy} 
\label{table:satellite_galaxy}
\begin{tabular}{ccc|cc|ccc}
\tableline
\multicolumn{3}{c|}{Stellar Disk} & \multicolumn{2}{c|}{Dark Matter}  &  \multicolumn{3}{c}{Gas Disk}\\
\tableline
$M_{*}$         &   $a_{*}$  &  $b_{*}$ &   $\rho_{d0}$        &   $r_{0}$  &   $M_{\rm gas}$   &   $a_{\rm gas}$   &   $b_{\rm gas}$ \\
($M_{\odot}$)   &	(kpc)    &	(kpc)	&	($\rm g~cm^{-3}$)  &   (kpc)	&	($M_{\odot}$)	&	(kpc)	&	(kpc)	\\
\tableline
$10^{9.7}$	&	2.5	&	0.5	&	5.93e-25	&	11.87   &   $10^{9.7}$	&	3.75	&	0.75 \\
\tableline
\end{tabular}
\end{table}

Among the three components in Table \ref{table:satellite_galaxy}, our simulations model the stellar disk and dark matter as static gravitational potential fields. The static stellar disk potential is under the Plummer-Kuzmin model \citep{miyamoto_three-dimensional_1975} with the scale length ($a_{*}$) and height ($b_{*}$) of $2.5$ and $0.5$ kpc, respectively (from the baryonic mass-stellar disk size scaling relation; \citealt{wu_scaling_2018}). We model the cold dark matter potential under the spherical Burkert model \citep{burkert_structure_1995,mori_gas_2000}, 
which is selected to better match the observational rotation curves of low-mass galaxies \citep{salucci_dark_2000,blok_high-resolution_2008}. Given the stellar mass (Table \ref{table:satellite_galaxy}), we obtain the circular velocity $V_{\rm circ} \approx 120$ $\rm km~s^{-1}$ from the observational baryonic Tully-Fisher relation \citep{lelli_baryonic_2019,mcgaugh_baryonic_2021}, which gives the dark matter central density $\rho_{d0}$ and scale radius $r_{0}$ (Table \ref{table:satellite_galaxy}). 

The gas disk in Table \ref{table:satellite_galaxy} is followed in our simulations with AMR. We adopt the gas mass from observed gas-(\HIspace and H$_{2}$ combined) to-stellar mass ratio $M_{\rm gas}/M_{*} \approx 1$ \citep{calette_hi-_2018}, and the disk size from the size ratio $R_{\rm gas}/R_{\rm optical} \approx 1.25$ \citep{swaters_westerbork_2002}. This ensures that the resulting galaxy model is consistent with the $z \approx 0$ observed scaling relations. The gas density is distributed under a softened exponential disk model \cite[see][eqn. 1]{tonnesen_gas_2009,tonnesen_tail_2010}, and the temperature and pressure are calculated to maintain hydrostatic equilibrium in the disk with the surrounding ICM. The rotational velocity is then calculated to balance the gravitational force and the combination of the centrifugal force and the pressure gradient.

\subsection{The orbits}\label{subsec:orbits}
We model the time-varying infalling orbits --- satellites travelling from the host's virial radius $R_{200}$ to the pericenter location $R_{\rm peri}$ --- of three host halos: a ``Milky Way-like" host halo of $M_{200} = 10^{12}$ $M_{\odot}$, a ``group" halo of $M_{200} = 10^{13}$ $M_{\odot}$, and a  ``cluster" halo of $M_{200} = 10^{14}$ $M_{\odot}$. The host mass selection is motivated by the mass-dependent quenched fraction disagreements in \cite{donnari_quenched_2021}: a satellite of $M_{*} = 10^{9.7}$ $M_{\odot}$ tends to be under-quenched in the TNG300 simulations compared with SDSS observations for the low-mass hosts $M_{200, \rm host}<10^{13.5}$ $M_{\odot}$, but over-quenched in simulations for the higher-mass hosts \cite[see][Fig 9]{donnari_quenched_2021}. 
Our host mass sampling ($M_{200, \rm host} \in [10^{12}, 10^{14}]$ $M_{\odot}$) is to span the mass range over which the \cite{donnari_quenched_2021} turnover in quenching effectiveness happens for the satellite of $M_{*} = 10^{9.7}$ $M_{\odot}$.

In this subsection, we describe our two-step orbit modeling process: (1) Satellite orbit kinematics (Table \ref{table:satellite_orbits}), which gives the position and velocity of the satellite galaxy as a function of infalling time; and (2) Host halo radial profiles (Table \ref{table:halo_profiles}), which gives the density and temperature of the host's gaseous halo medium as a function of radius.

\begin{deluxetable*}{cccccccccc}\label{table:satellite_orbits}
\tablecaption{Parameters of the Satellite Galaxy Infalling Orbits} 
\tablehead{\colhead{$M_{200}$} & \colhead{Case$^{(1)}$} & \colhead{$c^{(2)}$} & \colhead{$R_{200}^{(3)}$} & \colhead{$R_{\rm peri}^{(3)}$} &\colhead{$V_{200}^{(4)}$} & \colhead{$(|V_{\phi,0}|, |V_{r,0}|)^{(5)}/V_{200}$} & \colhead{$V_{\rm peri}^{(3)}$} & \colhead{$e^{(6)}$} & \colhead{$\tau_{\rm infall}^{(7)}$} \\
\colhead{($M_{\odot}$)} & \colhead{} & \colhead{} & \colhead{(kpc)} & \colhead{(kpc)} & \colhead{($\rm km~s^{-1}$)} & \colhead{} & \colhead{($\rm km~s^{-1}$)} & \colhead{} & \colhead{(Mpc)}}
\startdata
$10^{12}$	&	Milky-Way	&	8.81	&	211	&	75	&	143	&	(0.655, 0.832)	&	265	&	0.674	&	1127 \\
$10^{13}$	&	Group	&	7.08	&	455	&	149	&	308	&	(0.603, 0.786)	&	565	&	0.666	&	1165 \\
$10^{14}$	&	Cluster	&	5.62	&	949	&	278	&	663	&	(0.53, 0.782)	&	1236	&	0.692	&	1164 \\
\enddata
\tablecomments{$^{(1)}$ The case names represent the physical context of the central halos. $^{(2)}$ The present-day (redshift-zero) concentration values $c$ from \cite{ludlow_mass-concentration-redshift_2014}. $^{(3)}$ The virial radii ($R_{200}$), and the pericentric radii ($R_{\rm peri}$) and velocities ($V_{\rm peri}$) from the Gala-generated orbits, see \S \ref{subsec:orbits}. $^{(4)}$ The virial velocities defined as $V_{200} \equiv \sqrt{G \cdot M_{200}/R_{200}}$, following \cite{wetzel_orbits_2011}. $^{(5)}$ The tangential ($|V_{\phi,0}|$) and radial ($|V_{r,0}|$) velocity magnitudes at $R_{200}$ in units of the virial velocity ($V_{200}$; see note 4) from \cite{wetzel_orbits_2011}, used as velocity initial conditions for the orbit integration. $^{(6)}$ The resulting orbital eccentricities. $^{(7)}$ The infalling time from $R_{200}$ to $R_{\rm peri}$.}
\end{deluxetable*}

We use the Galactic Dynamics package Gala \citep{price-whelan_gala_2017,price-whelan_adrngala_2020} to perform time integration of the satellite orbits. First, we use Gala to construct the three host halos' gravitational potential profiles, adopting an NFW halo structure \citep{navarro_structure_1996}, and redshift-zero concentration values ($c$ in Table \ref{table:satellite_orbits}; \citealt{ludlow_mass-concentration-redshift_2014}). For simplicity, we assume that the satellite travels as a point mass when orbiting the host halos. The orbital integration begins at the host's virial radius $R_{200}$ (from the Gala-generated NFW profiles), and takes the best-fit values of \cite{wetzel_orbits_2011} as velocity initial conditions, see $(|V_{\phi,0}|, |V_{r,0}|)/V_{200}$ in Table \ref{table:satellite_orbits}. With the position and velocity initial inputs, we then use the Gala orbital integrator to integrate for a sufficient time (e.g., 100 Gyr) to ensure we capture many stable orbits, and focus on the branch from $R_{200}$ to pericenter $R_{\rm peri}$. The resulting orbits contain the satellite's position and velocity as a function of infalling time, as summarized in Table \ref{table:satellite_orbits}.

We model the extended, diffuse gaseous halos of the hosts as an isothermal sphere with a $\beta$-profile in density \citep{cavaliere_x-rays_1976,arnaud_-model_2009}. The spherical $\beta$ model is a relatively simple three-parameter model capable of reproducing the X-ray surface brightness observations for a range of galaxies \citep{makino_x-ray_1998,osullivan_x-ray_2003,anderson_detection_2011,dai_xmm-newton_2012}. It gives the gaseous halo density at a distance $r$ from the center as,
\begin{equation}\label{eqn:beta_profile}
    n(r) = n_{0} \cdot [1+(\frac{r}{r_{c}})^{2}]^{-3 \beta/2},
\end{equation}
where $n_{0}$ is the core density, $r_{c}$ is the core radius, and $\beta$ is the density slope at large radii. Among our three host cases, we apply the generic $\beta$-modeling for the group and cluster cases, but it breaks down at the low mass of $10^{12}$ $M_{\odot}$, where we instead use observational data of the Milky Way \citep{miller_structure_2013,miller_constraining_2015,salem_ram_2015,voit_ambient_2019}. The parameters of the halo models are summarized in Table \ref{table:halo_profiles} below.

\begin{deluxetable}{cccccc}\label{table:halo_profiles}
\tablecaption{Parameters of the Hosts' Gaseous Halo Profiles} 
\tablewidth{0pt}
\tablehead{\colhead{Case$^{(1)}$} & \colhead{$n_{0}r_{c}^{3\beta~(2)}$} & \colhead{$C^{(3)}$} & \colhead{$\beta$} & \colhead{$T$}   &   \colhead{Refs$^{(4)}$} \\
\colhead{} & \colhead{($10^{-2} \rm cm^{-3} kpc^{3\beta}$)} & \colhead{(kpc)} & \colhead{} & \colhead{(K)}  &   \colhead{}}
\startdata
Milky-Way &	1.35	&	2.73    &   0.5	&	2.51E6  &   MB15	\\ [0.5ex] 
\hline\hline
Case$^{(1)}$    &   $n_{0}$         &   $r_{c}$   &   $\beta$   &   $T$   &   Refs$^{(4)}$    \\[1ex] 
                &   ($\rm cm^{-3}$) &   (kpc)     &             &   (K)   &                   \\[1ex]
\hline
Group	&	0.0121	&   25  &	0.655	&	5.53e6	&   KS01    \\
Cluster	&   0.0071	&	76  &   0.675	&	2.02e7	&   KS01    \\
\enddata
\tablecomments{(1). Case names as in Table \ref{table:satellite_orbits}. We list the Milky Way case separately from the group and cluster cases because of the different methods; see \S \ref{subsec:orbits}. (2). Best fit parameter $n_{0}r_{c}^{3\beta}$ for the modified $\beta$ profile of \cite{miller_constraining_2015}, which fits $n_{0}$ and $r_{c}$ of a $\beta$ profile together (equations \ref{eqn:beta_profile} and \ref{eqn:modified_beta}). (3). The constant factor $C$ in equation \ref{eqn:modified_beta} to match with the LMC-constrained pericentric conditions of the Milky Way halo \citep{salem_ram_2015}. (4). References for our adopted $\beta$-profile (or modified in the Milky Way case) parameters, $\beta$, $r_{c}$, and the isothermal gas halo temperatures $T$. MB15: \cite{miller_constraining_2015}; KS01: \cite{komatsu_universal_2001}.}
\end{deluxetable}

For the ``Milky Way" host case, we refer to the \cite{miller_constraining_2015} parameterization of a modified $\beta$ profile, 
\begin{equation}\label{eqn:modified_beta}
    n(r) \approx \frac{n_{0}r_{c}^{3\beta}}{r^{3\beta}},
\end{equation}
where $n_{0}$, $r_{c}$, and $\beta$ are defined as in equation \ref{eqn:beta_profile} above \cite[see][eqn. 2]{miller_constraining_2015}.
However, this profile becomes less constrained at the large radii of our satellite's orbit (Table \ref{table:satellite_orbits}), and tends to underestimate densities compared with other studies \cite[see][Fig. 3]{voit_ambient_2019}. To address this density underestimation at large radii, we boost the best-fit \cite{miller_constraining_2015} model by a constant factor $C \approx 2.73$, obtained by matching the LMC-constrained pericentric (at 48.2 kpc) density from \cite{salem_ram_2015}.

For the galaxy group and cluster cases, we obtain the three parameters in the $\beta$ model (equation \ref{eqn:beta_profile}) as follows: We adopt $r_{c}$ and $\beta$ from the gas halo profiles of \cite{komatsu_universal_2001}, and solve for $n_{0}$ as an integration constant by assuming the gas-to-total mass fraction at $R_{500}$ is $\approx 10 \%$ (\citealt{lovisari_scaling_2015}; $f_{\rm gas,500}=M_{\rm gas,500}/M_{\rm tot,500} \approx 10\%$). The gas mass within $R_{500}$ under a $\beta$-profile in density (equation \ref{eqn:beta_profile}) can be written as,
\begin{equation}\label{eqn:gas_mass_integration}
\begin{split}
    M_{\rm gas,500} & = 4 \pi \rho_{\rm gas,0} \int_{0}^{R_{500}} [1+(\frac{r}{r_{c}})^{2}]^{-3 \beta/2} r^{2}  dr  \\
    & \approx 10 \% \cdot M_{\rm tot,500}
\end{split}
\end{equation}
where we supply $r_{c}$, $\beta$ from \cite{komatsu_universal_2001}, $M_{\rm tot,500}$ and $R_{500}$ from the Gala-generated host NFW halos, and the $f_{\rm gas,500}\approx 10\%$ relation from \cite{lovisari_scaling_2015}. Solving equation \ref{eqn:gas_mass_integration} for the integration constant $\rho_{\rm gas,0}$ gives the central mass density and hence the number density $n_{0}$.

\begin{figure}
    \centering
    \includegraphics[width=0.95\linewidth]{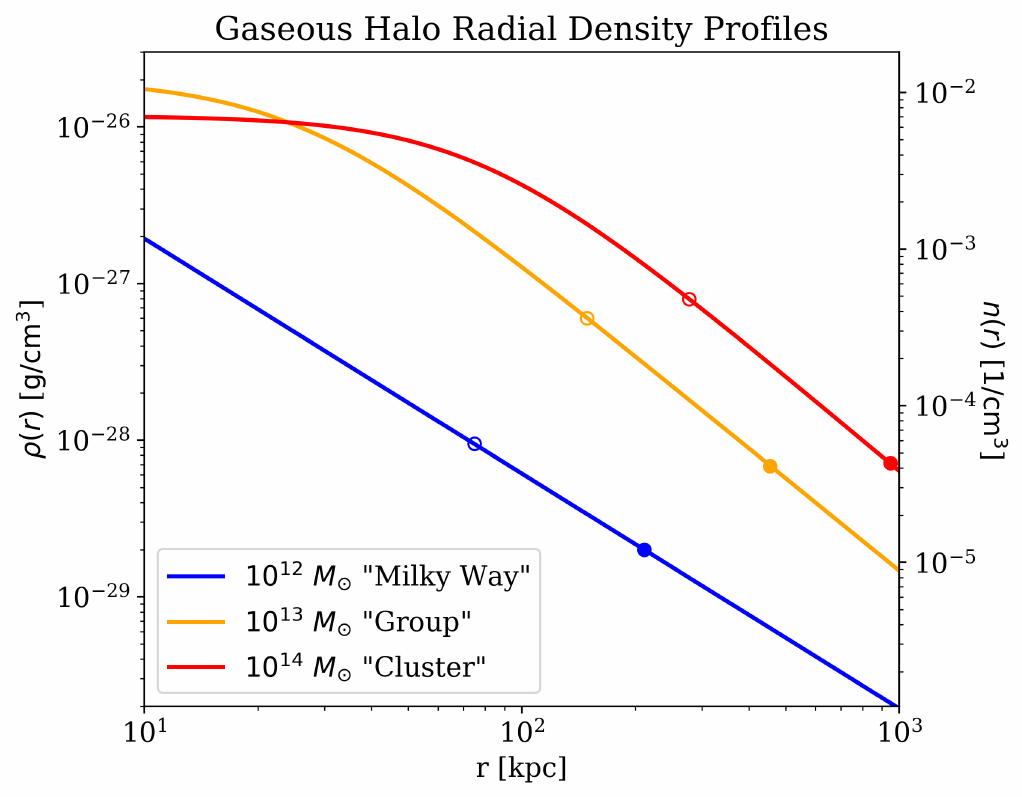}
    \caption{The radial density profiles of the three gaseous host halos in this study (Table \ref{table:halo_profiles}). Solid and empty circles show the virial radii ($R_{200}$) and pericenter radii ($R_{\rm peri}$), respectively, which mark the initial and final locations of the satellite infall orbits (see Table \ref{table:satellite_orbits}). The y-axis on each side shows the same information in mass density (left: $\rho$) and number density (right: $n$).}
    \label{fig:host_density}
\end{figure}

Figure \ref{fig:host_density} shows the density profiles of the three gaseous halo cases shown in Table \ref{table:halo_profiles}, where we annotate the infall orbits' initial (virial radius) and final (pericenter) locations from Table \ref{table:satellite_orbits}. At a given time $t$ of an infall orbit, the orbital density is given by the density profiles $\rho(r)$ in Figure \ref{fig:host_density}, taking the radius $r(t)$ from the Gala-generated orbits. The resulting orbital density ranges (densities between the solid and empty circles in Figure \ref{fig:host_density}) for the group and the cluster cases are relatively similar, but the cluster case has a higher pericentric velocity (Table \ref{table:satellite_orbits}), which leads to about five times the ram pressure of the group case at the pericenter, see Section \ref{subsec:simulations} below for details.

\subsection{The simulations}\label{subsec:simulations}
Our suite of four simulations includes three ``wind-tunnel" runs and one ``isolated galaxy" run. In each of the ``wind-tunnel" runs, we introduce a 45-degree-inclined boundary inflow (velocity normal vector $\hat{v}_{\rm wind(x,y,z)} = (0, \frac{\sqrt{2}}{2}, \frac{\sqrt{2}}{2})$) from the $y=0$, $z=0$ corner of the simulation box. Instead of a purely face-on or edge-on wind, we choose the 45$^{\circ}$ inclination angle to investigate the ram pressure effects both perpendicular and parallel to the disk. The inflow is modeling the ram pressure `wind', which carries the time-varying orbital conditions (gas density, temperature, and velocities) set in \S \ref{subsec:orbits}. The metallicities of the inflow gas (the `wind') and the initial galaxy gas disk (see \S \ref{subsec:satellite_galaxy}) are set as $Z_{\rm wind}=0.1 Z_{\odot}$, $Z_{\rm galaxy}=0.3 Z_{\odot}$, respectively, which are subsequently used as tracers for galactic versus wind material. 
The isolated galaxy run is a control case without inflow, but otherwise has the same setup of galaxy structure, radiative cooling, star formation, and feedback as in the wind runs.

\begin{figure*}
    \centering
    \includegraphics[width=0.95\linewidth]{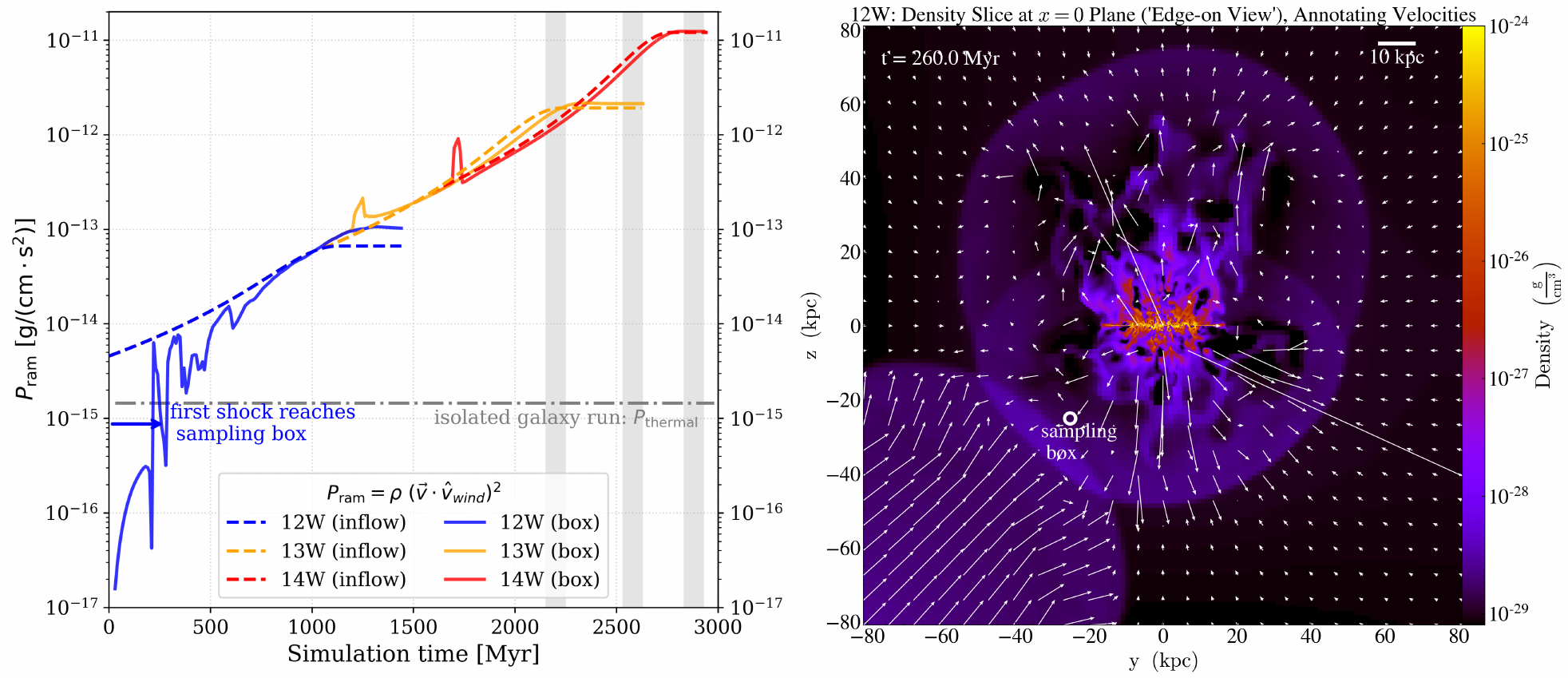}
    \caption{\textbf{Left}: Ram pressure time evolution for the three wind simulations. Dashed lines show the input from the orbits (\S \ref{subsec:orbits} and \S \ref{subsec:simulations}), solid lines are the simulation values ($\rho \cdot v_{\rm gas}^{2}$ in the wind direction $\hat{v}_{\rm wind}$) read in at a sampling box (see right panel). The dash-dotted gray line shows the ambient gas thermal pressure in the isolated run for reference. The three shaded vertical bars denote important time frames we will later refer to in Figure \ref{fig:gas_and_sfr_global} and \S \ref{sec:global_results}. \textbf{Right}: A density slice of the 12W ``Milky Way" case, annotating in-plane gas velocities. The inflow wind from the lower left corner travels at low velocities (Table \ref{table:satellite_orbits}), taking 200-300 Myr to reach the sampling box (white circle; at (0, -25, -25) kpc), and another 100-200 Myr to reach the disk (domain center). The inflow is about to reach the sampling box at this snapshot.}
    \label{fig:ram_pressure_evolution}
\end{figure*}

We summarize key aspects of the simulations in Table \ref{table:simulations} and Figure \ref{fig:ram_pressure_evolution}. The time-dependent ram pressure is defined as
$P_{\rm ram}(t) = \rho_{\rm host}(t) \cdot v_{\rm sat}(t)^{2}$, where the host halo medium density  (Figure \ref{fig:host_density}) $\rho_{\rm host}(t)$ is evaluated at the satellite location $r_{\rm sat}(t)$; and the satellite location and velocity, $r_{\rm sat}(t)$ and $v_{\rm sat}(t)$, are from the gala-generated orbits (\S \ref{subsec:orbits} and Table \ref{table:satellite_orbits}).
For the three wind runs (hereafter 12W, 13W, and 14W), we list the initial and final ram pressure values from the orbits described in \S \ref{subsec:orbits}, and show their time evolution in Figure \ref{fig:ram_pressure_evolution}. We initialize the 13W run from a snapshot in the 12W run where the ram pressure matches the 13W initial conditions, and similarly start the 14W run from a 13W snapshot; see the relevant time frames in Figure \ref{fig:ram_pressure_evolution}. This results in the initial galaxy disk in the 13W and 14W runs being ``pre-processed": it has been orbiting in smaller host halos prior to their accretion onto the more massive group- or cluster-host, a highly probable process for low-mass galaxies that has abundant observational and theoretical evidence \citep{zabludoff_environment_1996,wetzel_galaxy_2013,haines_locuss_2015,jung_origin_2018,bahe_disruption_2019,donnari_quenched_2021-1}. For the masses modeled in this paper, TNG simulations find $\approx 50 \%$ of satellites below $M_{*} = 10^{10}$ $M_{\odot}$ have been pre-processed in hosts of $M_{200}=10^{12}$ $M_{\odot}$ or above, if they reside in a cluster of $M_{200}=10^{14}$ $M_{\odot}$ at $z=0$ \citep{donnari_quenched_2021-1}. 

Initializing the simulations from a previous snapshot effectively avoids the numerical artifacts in the initial few hundred Myr, like an unstable outburst of star formation \citep{tasker_simulating_2006} and significant transient ringing \citep{goldbaum_mass_2015} in the gas disk. Previous works such as \cite{tonnesen_star_2012} addressed these artifacts by delaying the wind and allowing for a thermal relaxation phase of at least 200 Myr to stabilize the disk. For our 12W run (the only wind simulation that begins at $t=0$), however, the wind delay is unnecessary. Because of the Milky Way wind's initial slow speed ($|v_{\rm wind}|\approx 151$ $\rm km~s^{-1}$; Table \ref{table:satellite_orbits}), it takes the first shock wave (of Mach number 2) generated by the initial inflow more than 300 Myr to reach the galaxy disk, and longer for the stable inflow, see Figure \ref{fig:ram_pressure_evolution}. The location of the sampling box for ram pressure values (right panel of Figure \ref{fig:ram_pressure_evolution}) is chosen to be relatively close to the galactic disk, while avoiding the bow shock in front of the galaxy after the thermal relaxation phase.

The three wind runs cover over three orders of magnitude in ram pressure (solid lines in Figure \ref{fig:ram_pressure_evolution}), which generally follow the input orbit conditions (dashed lines). We attach a constant ram pressure value at the end of each wind run to ensure the pericentric inflow from the corner of the simulation box reaches the galactic disk, and the attached time periods are relatively short ($<$300 Myr). We annotate the input ICM thermal pressure of the isolated case as the dash-dotted line in Figure~\ref{fig:ram_pressure_evolution}, which is lower than the weakest ram pressure input of the wind runs ($t=0$ of 12W). 
The sampled ram pressure of 12W (blue solid line) is low during $0-300$ Myr because it shows the initial collapse of the gas before the wind reaches the sampling box; then during $\sim$$300-700$ Myr, its stochasticity reflects an interplay between the feedback outflows and ram pressure of a comparable strength. The two short peaks in ram pressure at $\sim$1250 (13W) and 1700 Myr (14W) are due to shock waves generated when we stack the ram pressure profiles; they have no global effects on the simulations.

\begin{deluxetable}{cccc}\label{table:simulations}
\tablecaption{Overview of the Simulation Suite} 
\decimalcolnumbers
\tablehead{\colhead{Simulation} & \colhead{$P_{\rm ram,i}$} & \colhead{$P_{\rm ram,f}$} & \colhead{$t_{i}$}\\
\colhead{} & \colhead{($\rm g/(cm \cdot s^{2})$)} & \colhead{($\rm g/(cm \cdot s^{2})$)} & \colhead{(Myr)}}
\startdata
12W	&	4.6e-15	&	6.7e-14	&	0\\
13W	&	6.4e-14	&	1.9e-12	&	1060\\
14W	&	2.6e-13	&	1.2e-11	&	1600\\
iso	&	\nodata	&	\nodata	&	0\\
\enddata
\tablecomments{(1) Simulation short names, used throughout this paper. 12W: ``Milky Way" halo ($10^{12}$ $M_{\odot}$) wind; 13W: ``group" halo ($10^{13}$ $M_{\odot}$) wind; and 14W: ``cluster" halo ($10^{14}$ $M_{\odot}$) wind; iso: isolated galaxy (no wind). (2) and (3) The initial and final ram pressure values of the wind orbits (\S \ref{subsec:orbits}), also see Figure \ref{fig:ram_pressure_evolution}. (4) The initial time of each simulation. 12W and iso both begin at $t=0$, but the 13W and 14W runs are each continued from a lower halo mass case's evolved snapshot with matching ram pressures. For example, the 13W run is a continuation from the $t=1060$ Myr snapshot of 12W, where the initial ram pressure $P_{\rm ram,i(13W)}$ matches the 12W run's $P_{\rm ram,t=1060(12W)}$, see Figure \ref{fig:ram_pressure_evolution}.}
\end{deluxetable}

\section{Global Results: Gas Stripping and Star Formation Rate Response}\label{sec:global_results}

We present our simulation results as follows: \S \ref{subsec:global_results} summarizes the global evolution of baryonic mass, star formation rate (SFR), and star forming location; \S \ref{subsec:gas_kinematics} describes the wind-driven gas morphology and kinematics, which explains the global evolution in \S \ref{subsec:global_results}.

\subsection{The Fate of the Ram Pressure Stripped Galaxy}\label{subsec:global_results}
\begin{figure}
    \centering
    \includegraphics[width=1.0\linewidth]{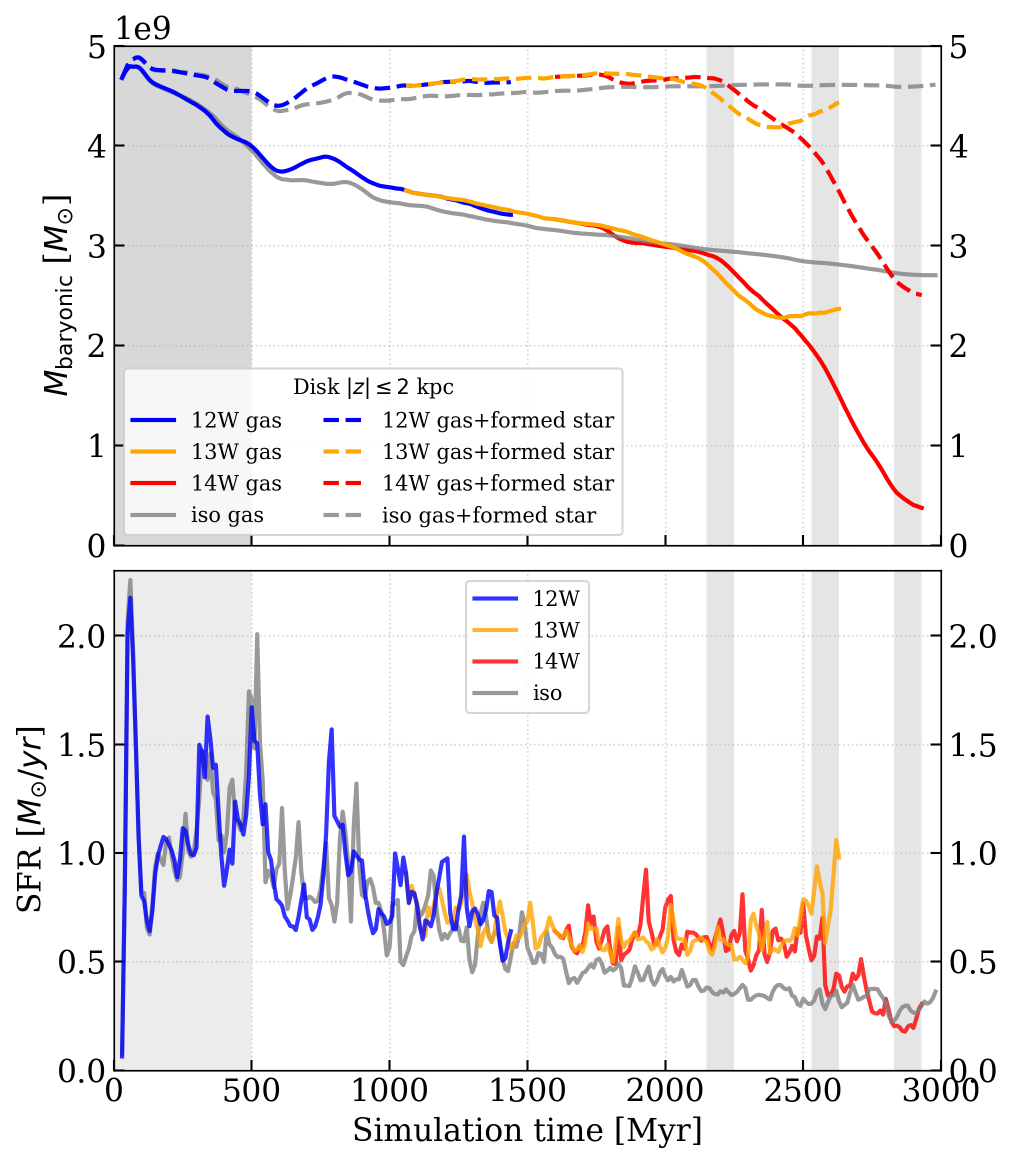}
    \caption{\textbf{Upper panel}: The time evolution of the galaxy disk's ISM mass (solid lines), where the ISM 
    is specified by a combined spatial and metallicity selection (\S \ref{subsec:global_results}).
    The dashed lines are the ISM mass in solid lines added with the total mass of formed stars under the same spatial selection. \textbf{Lower panel}: The SFR time evolution. The first $\sim$500 Myr is the thermal relaxation phase (\S \ref{subsec:simulations}), where the disk cools and collapses, leading to an initial burst in star formation, and gradually stabilizes. This unstable phase ($0-500$ Myr) is shaded and will be omitted in plots hereafter. To aid visual comparison, we denote three 100 Myr time periods with vertical bars showing the onset of effective stripping, 13W pericenter, and 14W pericenter from left to right; see \S \ref{subsec:global_results}.}
    \label{fig:gas_and_sfr_global}
\end{figure}

The global effects of RPS on the disk mass and SFR are demonstrated in Figure \ref{fig:gas_and_sfr_global}, comparing the three wind cases and the iso case. The upper panel shows the galactic disk masses: gas as solid lines and gas plus formed stars as dashed lines, versus simulation time. We apply a spatial disk cut ($R_{\rm disk} = 18$ kpc, $z_{\rm disk} = \pm 2$ kpc) and a metallicity cut ($Z_{\rm gas} \geq 0.25 Z_{\odot}$) to select the gas in the galactic disk, and place the same spatial cut when calculating masses of the formed stars. We verified that varying these selection criteria to include more gas (e.g., using $z_{\rm disk} = \pm 5$ kpc and $Z_{\rm gas} \geq 0.2 Z_{\odot}$) does not change the trends. The gas fuels star formation and decreases in mass in all four cases. Without RPS, star formation accounts for most of the gas mass loss, as shown by the iso case's gas plus formed stellar mass (dashed gray line), which remains at $\sim 98.5\%$ its initial value --- almost conserved after $\sim$3 Gyr of evolution over which time nearly $45\%$ of the initial gas mass is converted to stars.

In the wind runs, in addition to forming stars, the gas can gain or lose mass due to interactions with the wind, depending on the ram pressure strength. In 12W, there is no stripping-induced mass loss compared to the iso case due to the weak ram pressure (Figure~\ref{fig:ram_pressure_evolution}). Instead, there is a mild mass excess at $t \sim 750$ Myr, because ram pressure pushes part of the feedback outflows back to the disk (this will be discussed in \S \ref{subsec:gas_kinematics}). Both 13W and 14W experience the onset of the stripping phase at $t \sim 2150-2250$ Myr, seen from the steepened gas mass slopes in Figure~\ref{fig:gas_and_sfr_global} (leftmost vertical bar; see Figure \ref{fig:ram_pressure_evolution} for the corresponding $P_{\rm ram}$). The stripping only lasts for $\sim$200 Myr in 13W, after which the gas mass slope flattens 
as the ram pressure is kept constant (Figure~\ref{fig:ram_pressure_evolution}). But in 14W, as the ram pressure continues to increase, the stripping continues until nearly the entire gas disk is removed.

The lower panel of Figure \ref{fig:gas_and_sfr_global} shows the SFR evolution, manifesting both the SFR enhancing and quenching potential of the wind. The first $\sim$500 Myr is the thermal relaxation phase where star formation is still stabilizing, and the 12W wind has yet to reach the galaxy (Figure~\ref{fig:ram_pressure_evolution} and \S \ref{subsec:simulations}). This period will be omitted in subsequent plots and analyses. After the relaxation phase, the SFR steadily decreases in the iso case throughout the $\sim$3 Gyr of evolution, as its gas density steadily decreases due to starvation without cosmological inflow replenishing the disk. The SFR of 12W remains similar to the iso case throughout its orbit. In 13W and 14W, the SFR remains approximately constant at $\sim 0.6$ $M_{\odot}/\rm yr$ until $t \sim 2450$ Myr, which is a relative enhancement compared with iso. The SFR then mildly increases in 13W as the gas mass remains almost constant, resulting in a 2.5 times SFR enhancement relative to iso at the first-infall pericenter (middle vertical bar). In 14W during the final $480$ Myr, the SFR decreases by $\sim 65\%$, dropping to below iso at the cluster pericenter (rightmost vertical bar), as the gas is rapidly removed from the disk (from $\sim 2.2\times 10^{9}$ $M_{\odot}$ to $\sim 3.8\times 10^{8}$ $M_{\odot}$). The 14W galaxy will ultimately be quenched judging from the rapid, almost complete gas removal.

Combining the ram pressure, the disk mass, and SFR (Figures \ref{fig:ram_pressure_evolution} and \ref{fig:gas_and_sfr_global}) gives the direct effect of ram pressure on these star-forming disks' global evolution. The turning point at $t \sim 2200$ Myr, where the effective gas stripping takes place in 13W and 14W, corresponds to a total 45$^{\circ}$-angled ram pressure of $P_{\rm ram} \approx 1.5 \times 10^{-12}$ g/($\mathrm{cm} \cdot \mathrm{s}^{2}$). 
Before this critical ram pressure is reached, the gas plus formed stellar masses in the disks (dashed lines in Figure \ref{fig:gas_and_sfr_global}) remain conserved with respect to ram pressure for all wind runs. Unlike gas stripping, which is a direct consequence of strong ram pressure, the SFR shows no immediate correlation to ram pressure. The SFR turning points in 13W and 14W appear $\sim$300 Myr delayed compared to the gas mass change (Figure \ref{fig:gas_and_sfr_global}), likely because it takes time for the wind-driven mass flows around the disk to affect the global SFR.

\begin{figure}
    \centering
    \includegraphics[width=1.0\linewidth]{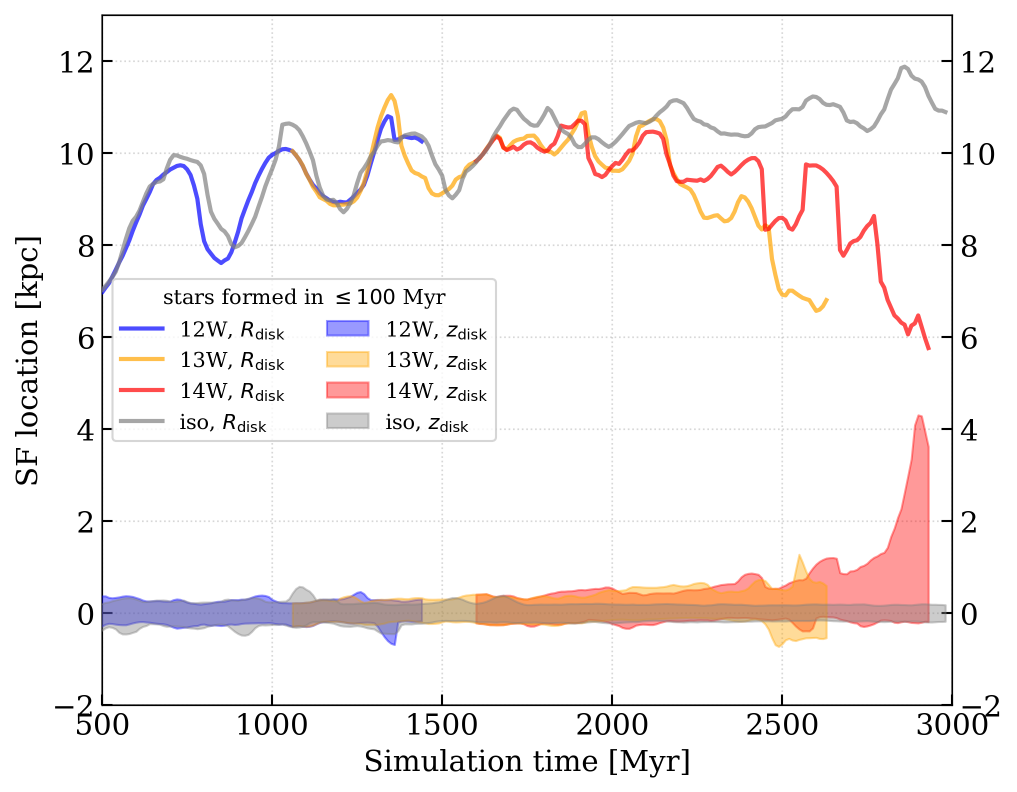}
    \caption{The evolution of star forming locations in terms of the cylindrical/disk radius $R_{\rm disk}$ and height $z_{\rm disk}$ versus simulation time. The radii shown here as solid lines are the $95$ percentile values, and the heights as colored bands are $5-95$ percentile values. We select new stars formed within 100 Myr to match the typical timescales of UV observations (\S \ref{subsec:global_results}). Higher ram pressures lead to more central star formation.}
    \label{fig:sf_radius_and_height}
\end{figure}

As the wind interacts with the galactic disk, the location of star formation changes, as shown in Figure~\ref{fig:sf_radius_and_height}. We select stars newly formed within 100 Myr of each given time and obtain their distribution in cylindrical/disk radius ($R_{\rm disk}$) and height ($z_{\rm disk}$). These stars' radial distribution has a power-law tail, and Figure \ref{fig:sf_radius_and_height} shows their $95\%$ percentile values as solid lines. The height distribution characterizes a thin disk of cold gas --- symmetrically peaked around $z=0$ kpc without wind impact, or skewed towards $+z$ under a wind that has a $+z$ component, and Figure~\ref{fig:sf_radius_and_height} shows their $5 - 95\%$ percentiles as colored bands. 

The radial ringing in Figure \ref{fig:sf_radius_and_height} results from epicyclic oscillations triggered by rapid radiative cooling disturbing the initial equilibrium disk \citep{goldbaum_mass_2015}, and has no global effect on our results. In the iso case, the radius of star formation in an undisturbed disk shows a slow and steady increase and asymptotes to $\sim$11 kpc at the final stage of evolution, and the height remains symmetric within $\pm 0.5$ kpc, which 12W closely follows throughout its orbit. In 13W and 14W, when the gas stripping begins at $t \sim 2200$ Myr (Figure \ref{fig:gas_and_sfr_global}), their radii of star formation begin to deviate from the iso case, with the 13W radius decreasing slightly faster and its $z_{\rm disk}$ symmetrically thickening, while the 14W radius decreasing relatively slowly to $\sim$6 kpc and its $z_{\rm disk}$ extending to $+4$ kpc, highly skewed towards the wind direction. The star formation radius in 13W decreases faster than in 14W, where ram pressure is higher, which seems to contradict the \cite{gunn_infall_1972} face-on stripping picture. But this is because of the wind inclination: more 14W gas is stripped into an extensive tail inclined to the disk (see Figure~\ref{fig:mass_flow_via_density_slice} below), forming 
stars 
in the tail, which skews the 14W star formation to higher cylindrical radii.

We selected the 100 Myr timescale in Figure \ref{fig:sf_radius_and_height} in order to match the typical timescales in UV observations of star formation (e.g., \citealt{leroy_estimating_2012,kennicutt_star_2012}). We also experimented with 10 Myr (typical H$\alpha$ timescale), 30 Myr, and all formed stars within the simulations (1-3 Gyr, roughly matching the timescales in optical observations, see \citealt{tasker_simulating_2006}), and found the temporal trends agree for the 10, 30, 100 Myr selections. If using all formed stars, however, the radial distributions in all cases asymptote to $R_{\rm disk} \approx 10-11$ kpc, characterizing a steady stellar disk (rather than recent star formation); the similar radius across all cases is expected as RPS has no direct effect on the stellar disk.

\subsection{Wind-driven Gas Morphology and Kinematics}\label{subsec:gas_kinematics}

The global results in \S \ref{subsec:global_results} show that RPS directly affects the gas mass and, although the global SFR is eventually affected, that impact occurs a few hundred Myr after the onset of gas stripping. We find the impact on star formation can evolve in opposite directions: enhance or quench (Figure \ref{fig:gas_and_sfr_global}). In this section, we examine the wind-driven gas flows to determine the physical reasons behind the bimodal effects on star formation. 

\begin{figure*}
    \centering
    \includegraphics[width=1.0\linewidth]{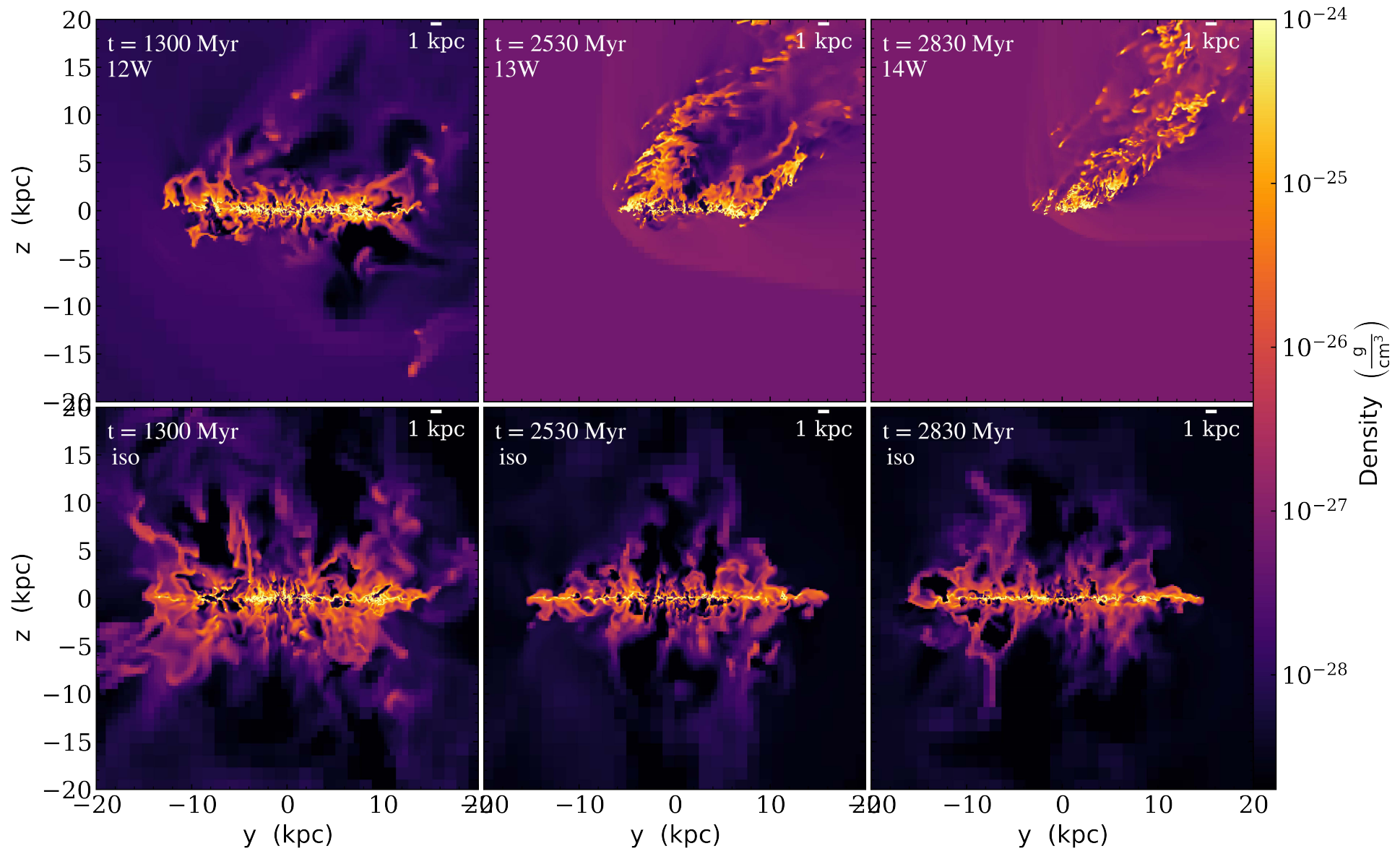}
    \caption{Gas density ``edge-on" slices, zoomed in to $40$ kpc. The 45$^{\circ}$ winds introduced as boundary inflows are in the $\hat{v}_{\rm wind(y,z)} = (\frac{\sqrt{2}}{2}, \frac{\sqrt{2}}{2})$ direction (\S \ref{subsec:simulations}). The selected time frames, as annotated on the upper left corners, correspond to the pericentric conditions of the 12W, 13W, and 14W orbits (Figures \ref{fig:ram_pressure_evolution} and \ref{fig:gas_and_sfr_global}). The upper panels show the wind cases, and the lower panels show the isolated case at the same time for comparison.}
    \label{fig:mass_flow_via_density_slice}
\end{figure*}

We first show the different morphology of the gas via density slices along the $x = 0$ plane in Figure \ref{fig:mass_flow_via_density_slice}, comparing the iso case to each wind run when the ram pressure has 
reached its peak value at the galaxy position. The three lower panels show that, without RPS, the isolated galactic disk remains cylindrically symmetric and drives an outflow above and below the disk via star formation feedback, which decays in strength as the SFR decreases with time (Figure \ref{fig:gas_and_sfr_global}). The three wind runs demonstrate different interactions between the wind and the galactic gas, as summarized below.

\begin{itemize}[nolistsep]
    \item 12W: 
    There is no clear signal of RPS within the gas disk. Ram pressure appears to interact with the feedback-driven outflows, likely suppressing those outflows below the disk (against the wind direction).
    
    \item 13W: Gas is being stripped and forms an outer ring 
    that, in this slice, looks like two tails. The feedback-launched gas below the disk in the iso and 12W runs is missing because of the higher ram pressure.
    
    \item 14W: Gas is being stripped relatively uniformly from all radii of a shrunken and highly fragmented disk, forming a single extended tail tracing the wind direction. The 14W wind has a similar density but $>$2 times higher velocity compared with 13W, leading to its $>$4 times higher pericentric ram pressure (Tables \ref{table:satellite_orbits} and \ref{table:simulations}).
    
\end{itemize}

\begin{figure*}
    \centering
    \includegraphics[width=1.0\linewidth]{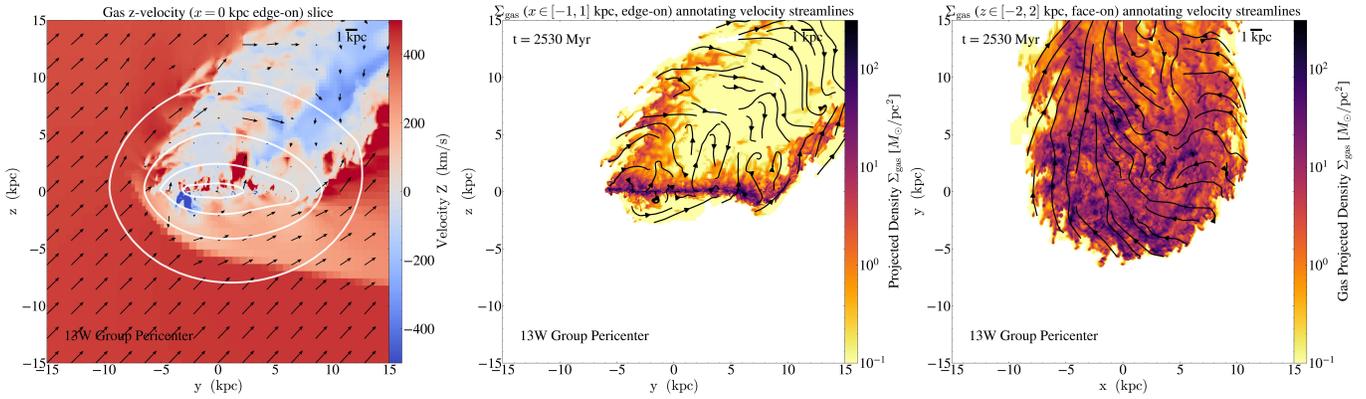}
    \caption{Gas z-velocity slice and projected density maps of 13W at the pericenter (as in Figure \ref{fig:mass_flow_via_density_slice}), zoomed in to $30$ kpc. \textbf{Left panel}: Gas $v_{z}$ slice map, annotating gravitational potential contours in white lines. The potential includes the static stellar and dark matter components (\S \ref{subsec:satellite_galaxy}), self-gravity of the gas, and of the newly formed stars. The in-plane velocity vectors ($v_{y,z}$) are marked by black arrows. \textbf{Middle and right panels}: Gas density projection maps (edge-on and face-on) with a metallicity selection for the galactic ISM ($Z_{\rm gas} \geq 0.25$ $Z_{\odot}$; largely unmixed with the ICM). The streamlines show the mass-weighted gas velocities.}
    \label{fig:grav_potential_fallback_and_mixing}
\end{figure*}

Moving from 12W to 13W and finally to 14W, we see a clear progression from negligible stripping to outer gas removal to nearly complete stripping. However, 
the 13W gas morphology in Figure \ref{fig:mass_flow_via_density_slice} demonstrates a unique complexity: within the stripping radius (characterized by the outer ring of stripped gas), there is high-density gas above the disk. 
We zoom in and examine the gas kinematics of 13W in Figure \ref{fig:grav_potential_fallback_and_mixing}. The $v_{z}$ slice (left; not mass-weighted) shows that much of the gas above the disk has a $\sim$0 or negative z-velocity --- falling back to the disk. The stripped material from the leading side ($y<0$), initially traveling at 45$^{\circ}$ ($\hat{v}_{\rm wind}$), experiences gravitational forces toward the disk center where the potential well is deepest (perpendicular to the contour lines, Figure~\ref{fig:grav_potential_fallback_and_mixing}). This fallback phenomenon is confirmed by the mass-weighted gas velocity streamlines, indicating the paths of motion, in the edge-on projection map of gas that originated in the galaxy (middle panel). Fallback happens for the relatively dense stripped gas at the wind leading edge ($y \approx -6$ kpc) onto the leading half of the disk, and also for the more diffuse stripped gas in the tail ($z \gtrsim 5$ kpc), which occurs on a larger spatial scale.

The face-on projection map (Figure \ref{fig:grav_potential_fallback_and_mixing} right panel), on the other hand, shows the interplay between ram pressure and disk rotation within the disk plane. On the $x>0$ side of the disk where the ram pressure in-plane component ($+y$) counters disk rotation, the disk gas can lose its angular momentum, manifested in radial inflows towards the disk center. This is clearly distinct from the $x<0$ side, where ram pressure aligns with disk rotation, and the in-plane gas kinematics transforms into radial outflows in the wind-trailing end of the disk ($y>0$).  Gas that is pushed above the disk while losing angular momentum, as illustrated in the third panel, will be more able to fall back along the streamlines shown in the middle panel.

\begin{figure*}[!htb]
    \centering
    \includegraphics[width=0.9\linewidth]{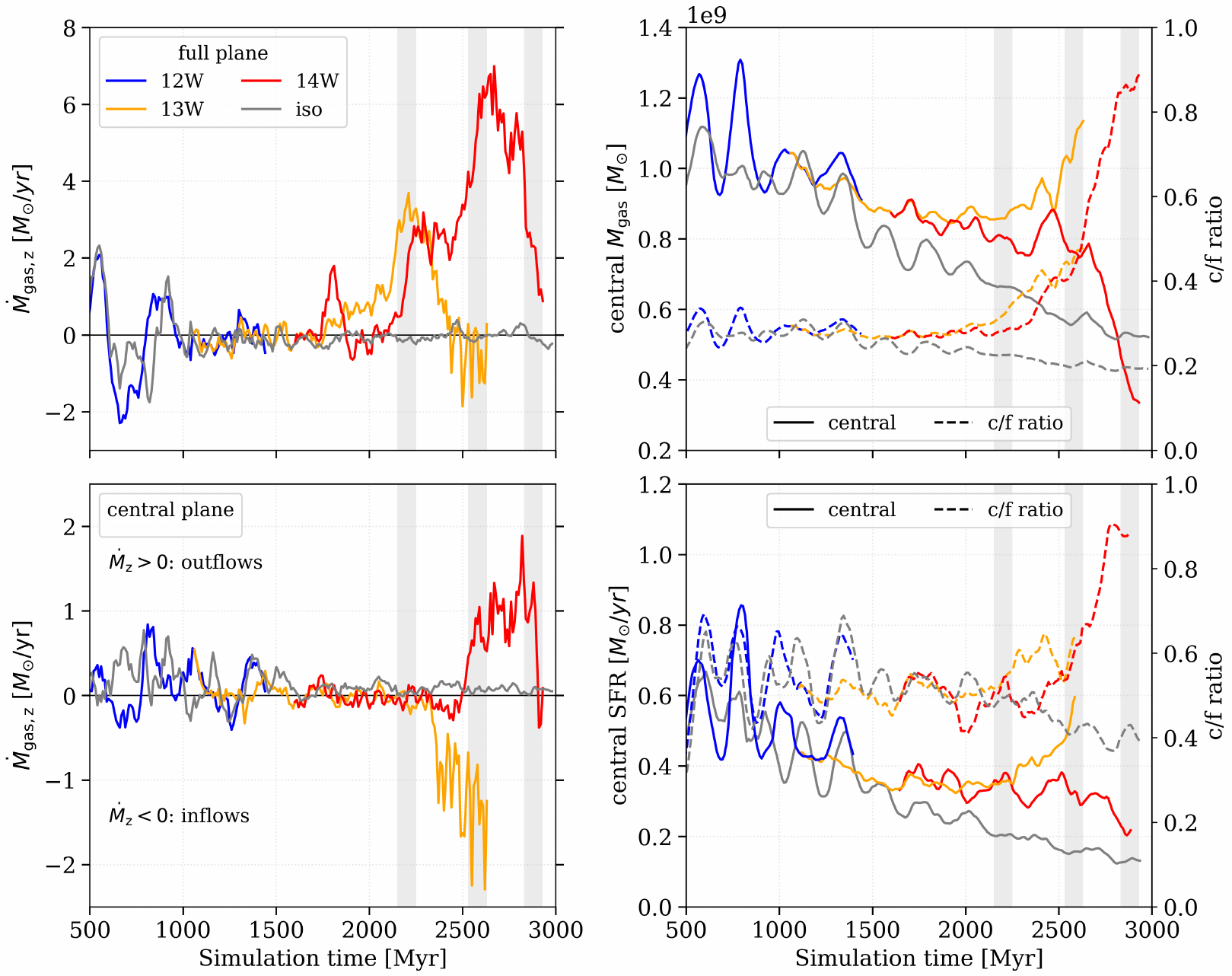}
    \caption{\textbf{Left}: Gas mass loss rate perpendicular to the disk ($\dot{M}_{\rm gas,z}$, where positive values indicate outflows), evaluated at disk heights $z = \pm 2$ kpc, for the full plane (upper left) and the central 5 kpc regions (lower left). \textbf{Right}: The central 5 kpc gas mass (upper right) and SFR (lower right) time evolution as solid lines (similar to Figure~\ref{fig:gas_and_sfr_global}, here for the central disk), and the central-to-full disk ratio (``c/f ratio") as dashed lines on the right-hand y axes. In all panels, we show the three 100 Myr reference time frames as in Figure \ref{fig:gas_and_sfr_global}.}
    \label{fig:mass_flow_mdot_series}
\end{figure*}

We now focus on RPS-driven mass losses using the gas motions perpendicular to the disk 
\citep{roediger_ram_2006,bekki_galactic_2014}. In our simulations, this corresponds to flows in the z-direction, $\dot{M}_{z} = \int_{\rm surface} \rho~\vec{v} \cdot d\vec{A} = \int_{\rm surface} \rho~v_{z} \cdot dA_{(x,y)}$, where the surfaces are selected to be $z_{\rm disk} = \pm 2$ kpc for the gas with metallicity $Z \geq 0.25 Z_{\odot}$\footnote{The metallicity selection excludes the ICM inflows (\S \ref{subsec:simulations}).}, consistent with our disk ISM selection (Figure \ref{fig:gas_and_sfr_global}). In addition, the mass flows are expected to have radial dependence, because gas removal typically begins at larger disk radii where the local gravity is weakest, and migrates radially inward as the ram pressure increases (\citealt{gunn_infall_1972}, also see Figure \ref{fig:sf_radius_and_height}). To characterize the radial dependence, we further distinguish the mass flows across the full planes ($z_{\rm disk} = \pm 2$ kpc) versus only the central 5 kpc regions of the planes ($R_{\rm disk} \leq 5$ kpc, $z_{\rm disk} = \pm 2$ kpc). The resulting $\dot{M}_{z}$ of the full and central planes are shown in Figure \ref{fig:mass_flow_mdot_series}, together with the gas mass and SFR time evolution (similar to Figure \ref{fig:gas_and_sfr_global}) of the central plane.

The mass loss rates in Figure \ref{fig:mass_flow_mdot_series}'s left two panels are obtained by summing $\dot{M}_{\rm gas,z}$ above and below the disk (at $z=+2$ and $-2$ kpc), where a positive value is outflow/mass loss by definition. For the iso case, the mass loss rate demonstrates galactic fountain flows driven by the stochastic star formation feedback alone: in the central plane where the star formation and feedback is strongest, $\dot{M}_{\rm gas,z}$ remains an outflow (lower left panel); but the full plane $\dot{M}_{\rm gas,z}$ oscillates around 0 (upper left panel), and results in $\sim$0 net baryonic mass loss throughout the 3 Gyr iso simulation (gray dashed line, Figure~\ref{fig:gas_and_sfr_global}). The fountain flows' (or central plane feedback outflows') amplitudes decay as the SFR decreases with time (Figure \ref{fig:gas_and_sfr_global}). 
The mass loss rate of 12W overall follows iso, except for a mildly enhanced inflow within its full plane (upper left panel) at $t \sim 600-800$ Myr, which explains the mild 12W gas mass excess around this time (Figure \ref{fig:gas_and_sfr_global}). We verified that the 12W enhanced inflow relative to iso is via the $z=-2$ kpc surface, indicating that the ram pressure, although not yet sufficient to strip the gas disk, transfers momentum with the diffuse fountain flows below the disk. 

For 13W and 14W, the mass loss rates are dominated by RPS. Across the full plane, the first peak of mass loss occurs at $t \sim 2150-2250$ Myr (leftmost vertical bar, Figure \ref{fig:mass_flow_mdot_series}), corresponding to the onset of effective stripping in both 13W and 14W (Figure \ref{fig:gas_and_sfr_global}). After that, the 13W mass loss rate steadily decreases to $\sim$0 at its pericenter, as the ram pressure becomes constant (Figure \ref{fig:ram_pressure_evolution}), and the 14W mass loss rate keeps increasing for another $\sim$400 Myr with its still increasing ram pressure. For the central plane, however, there is a clear dichotomy: 13W shows a central inflow ($\dot{M}_{\rm gas,z} < 0$) with increasing amplitude, while 14W shows a (slightly delayed) central outflow. During the onset of effective stripping, the stripping radius is greater than the selected central region (5 kpc), so $\dot{M}_{\rm gas,z} \sim 0$ for both cases. At $t \sim 2500$ Myr, the 14W stripping radius reaches the inner 5 kpc, and hence the central outflows begin. But for 13W, ram pressure was never sufficient to strip the inner disk; instead, the gravitational fallback of the stripped material (as described in Figure \ref{fig:grav_potential_fallback_and_mixing}) replenishes the central disk. 

The right-hand two panels of Figure \ref{fig:mass_flow_mdot_series} show the central disk gas mass and SFR time evolution and the central-to-full disk ratios (``c/f ratio"; see Figure \ref{fig:gas_and_sfr_global} for the full disk). For all four simulations, the central gas mass evolution tightly correlates with the central SFR evolution. The temporal oscillations arise from the radial ringing discussed previously in Figure \ref{fig:sf_radius_and_height}. In the absence of effective RPS, the iso and 12W cases maintain an almost constant c/f ratio throughout the simulations (gray and blue dashed lines). Effective RPS (13W and 14W) leads to a radial redistribution of the gas and SFR: (i) the profiles are more radially centralized (enhanced c/f ratios), and (ii) the inner disk $\rm M_{gas}$ and SFR values (solid lines) are both enhanced relative to iso. Although (i) partially results from the removal of outer disk gas, (ii) directly reflects the star formation enhancement potential of RPS.

\begin{figure*}[!ht]
    \centering
    \includegraphics[width=0.9\linewidth]{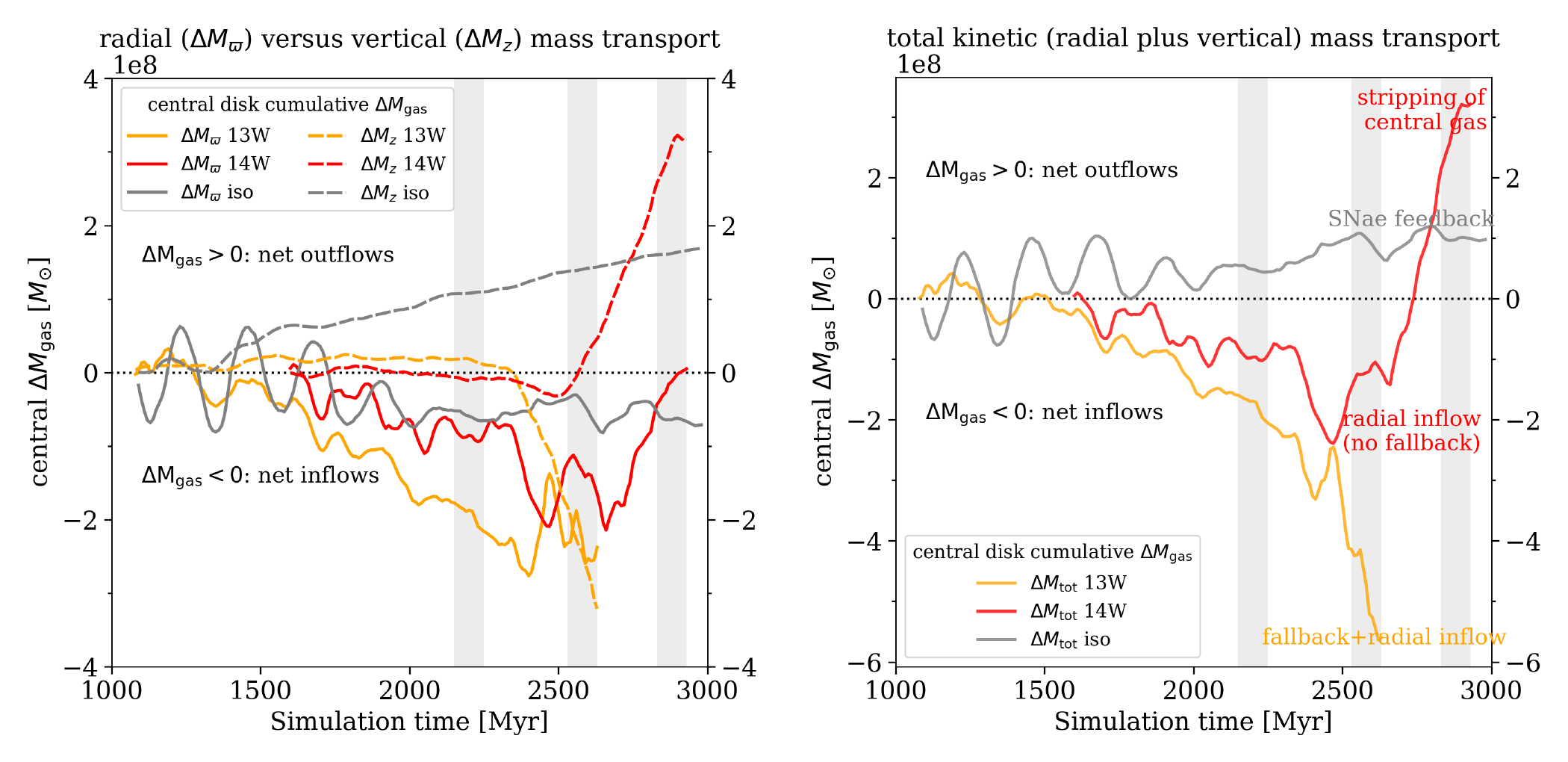}
    \caption{The cumulative mass loss ($\Delta \rm M_{gas}$) over time for the central 5 kpc disk as driven by mass transport, comparing 13W, 14W, and iso. Left: the cylindrical radial ($\Delta M_{\varpi}$; solid lines) versus vertical ($\Delta M_{z}$; see Figure \ref{fig:mass_flow_mdot_series} for $\dot{M}_{z}$, here dashed lines) mass loss components. Right: the total kinetic mass loss by summing the two components. As in Figure \ref{fig:mass_flow_mdot_series}, positive $\Delta \rm M_{gas}$ values denote outflows, and the vertical bars show the three reference time frames. Key physical processes that explain the trends are annotated on the right panel.}
    \label{fig:cumulative_central_flows}
\end{figure*}

The gas motions perpendicular to the disk (Figure \ref{fig:mass_flow_mdot_series}) demonstrated an indirect mode of radial mass transfer in 13W: gas is lifted by RPS from the edge of the disk, and falls back to the central disk a few 100 Myr later, replenishing star formation there (also see Figure \ref{fig:grav_potential_fallback_and_mixing}, middle panel). Another mode of radial mass transfer is directly via the (cylindrical) radial direction, $\dot{M_{\varpi}} = \int_{\rm surface} \rho~\vec{v} \cdot d\vec{A} = \int_{\rm surface} \rho~v_{\varpi}~dA$, where $\hat{\varpi}$ is the cylindrical radial vector, and the surface can be approximated by a thin cylindrical shell of average width $\bar{h}$, such that $\dot{M}_{\varpi} \approx (1/\bar{h})~ \sum_{i}^{\rm shell} (m_{i}~v_{\varpi,i})$. We evaluated the radial mass flow rate for the central disk in Figure \ref{fig:cumulative_central_flows} ($R_{\rm disk}\leq 5$ kpc, $|z_{\rm disk}| \leq 2$ kpc)\footnote{The shell width $\bar{h}$ is obtained by $\bar{h}= (1/A_{\rm shell}) \sum_{i}^{\rm shell} V_{i}$, where $V_{i}$ is the individual cell volume, and $A_{\rm shell}=2 \pi R_{\rm disk} \cdot (2 z_{\rm disk})$ is the shell area. We tested a range of shell widths, $\bar{h} \in [39,156]$ pc, which corresponds to 1 to 4 times the highest-refined cell length (\S \ref{sec:methods}), and $\dot{M}_{\varpi}$ is approximately constant over these $\bar{h}$. After obtaining $\bar{h}$, we evaluate $\dot{M}_{\varpi}(\bar{h})$ for the cells in the thin shell that satisfy the ISM metallicity selection $Z \geq 0.25 Z_{\odot}$. The final $\dot{M}_{\varpi}$ is an average of $\dot{M}_{\varpi}(\bar{h})$ over $\bar{h} \in [39,156]$ pc.}. Because $\dot{M_{\varpi}}$ characterizes mass transfer within the galaxy, its amplitude is particularly susceptible to radial oscillations (Figure \ref{fig:sf_radius_and_height}). Therefore, we show the time cumulative ($\Delta M = \sum \dot{M}(t)\Delta t$; frequent temporal oscillations cancel out) mass flows in Figure \ref{fig:cumulative_central_flows}: the left panel compares the radial and vertical (perpendicular to the disk) components, the right shows the total kinetic flows (the sum of the two).

The dashed lines in Figure \ref{fig:cumulative_central_flows}'s left panel is the time integration of the central plane $\dot{M}_{\rm gas,z}$ (Figure \ref{fig:mass_flow_mdot_series} lower left panel). As described above, in the central plane there is a consistent feedback outflow in iso, fallback replenishment in 13W, and stripping in 14W when $P_{\rm ram}$ becomes sufficient to affect the inner disk ($R_{\rm disk}\leq 5$ kpc). In the radial direction (solid lines, Figure \ref{fig:cumulative_central_flows} left panel), however, both 13W and 14W show an excess of inflow relative to iso, with peak amplitudes ($\Delta M_{\rm inflow} \approx 2 \times 10^{8}$ $M_{\odot}$) comparable to the net fallback inflow in 13W. These radial inflows can be explained by an interplay between the edge-on ram pressure component and disk rotation (right panel of Figure \ref{fig:grav_potential_fallback_and_mixing}). Rotating gas in the disk, when countered by ram pressure (where $x>0$; Figure \ref{fig:grav_potential_fallback_and_mixing}), can lose angular momentum and migrate radially inward. For 14W, this radial inflow eventually decreases to 0 as stripping proceeds into the central region of the disk with increasing $P_{\rm ram}$.

The right panel of Figure \ref{fig:cumulative_central_flows} shows the sum of the radial and vertical components: the total kinetic mass transport across all surfaces of the central disk. We summarize the key physical processes at play in the figure. For iso, the gradual mass loss is dominated by the central plane's star formation feedback. For 13W, there is a combination of replenishment from fallback and from direct radial inflow, resulting in its highest central plane gas mass and SFR (Figure \ref{fig:mass_flow_mdot_series} right panels). For 14W, because the ram pressure keeps increasing, there is net replenishment from radial inflow but little to no fallback around the group pericentric time (middle vertical bar), which eventually becomes a net outflow as the stripping radius reaches within the selected central disk ($R_{\rm disk}=5$ kpc) at the cluster pericenter (rightmost vertical bar). Collectively, the kinetic mass transport (Figure \ref{fig:cumulative_central_flows}) explains the central plane $M_{\rm gas}$ and SFR evolution (Figure \ref{fig:mass_flow_mdot_series}). We found RPS-driven direct radial inflows, in agreement with literature results \citep{schulz_multi_2001,tonnesen_gas_2009,akerman_how_2023}, can replenish the central star-forming disk; and we identified fallback as an indirect mode of radial mass transport (Figures \ref{fig:grav_potential_fallback_and_mixing}, \ref{fig:mass_flow_mdot_series}, and \ref{fig:cumulative_central_flows}) that can add to the enhancement for certain orbits.

We emphasize that as long as the $P_{\rm ram}(t)$ profiles are consistent, even with different $\rho_{\rm ICM}$, $v_{\rm sat}$ components (\S \ref{subsec:simulations}), the galaxy undergoes a similar global evolution. For example, at $t \sim 2200$ Myr (leftmost vertical bar), the global properties and the mass loss rates of 13W and 14W closely match (Figures \ref{fig:gas_and_sfr_global}, \ref{fig:mass_flow_mdot_series}, and \ref{fig:cumulative_central_flows}) as their $P_{\rm ram}$ values are comparable (Figure \ref{fig:ram_pressure_evolution}), despite the 14W orbit consisting of a higher $v_{\rm sat}$ and a lower $\rho_{\rm ICM}$. Importantly, the galaxy's global evolution is also sensitive to the time derivative of ram pressure, $d P_{\rm ram}/dt$. When the 13W ram pressure stops increasing as the galaxy reaches the group pericenter ($d P_{\rm ram}/dt=0$, Figure~\ref{fig:ram_pressure_evolution} leftmost to middle vertical bar), the stripping radius is kept at constant and the remaining gas disk acts as a ``shield" for the stripped gas above it, such that gravity outweighs ram pressure in shielded regions and causes the central plane fallback ($\dot{M}_{z}$, Figure \ref{fig:mass_flow_mdot_series}). Conversely, in 14W where $d P_{\rm ram}/dt$ keeps increasing after $t \sim 2200$ Myr for $\sim$500 Myr, the stripping radius decreases, so the shielded region shrinks (Figure \ref{fig:mass_flow_via_density_slice}); without shielding the stripped gas above the disk is unable to fall back.

The global evolution described in \S \ref{subsec:global_results} and \ref{subsec:gas_kinematics} can be summarized as follows. In the iso run, the gas disk primarily loses mass to steady star formation, which drives feedback fountain flows that decrease in magnitude with decreasing SFR. In 12W with weak $P_{\rm ram}$, the global properties are overall consistent with iso, other than the additional interactions between the wind and the low-density fountain flows. In 13W with moderate $P_{\rm ram}$, RPS in the disk outskirts dominates the mass loss, and actually enhances the SFR in the remaining disk. Because the pericentric $P_{\rm ram}$ is insufficient to remove the entire gas disk, gas can migrate radially inward via fallback and direct radial inflows, both replenishing the central disk's star formation. 
In 14W, $P_{\rm ram}$ is sufficient to strip the gas first in the outskirts and then in the center, ultimately resulting in a rapid decline in the galaxy's SFR.

\section{Spatially Resolved Star Formation Rate-Mass Relation}\label{sec:resolved_SF_mass}

In the previous section, we found that RPS can enhance the satellite galaxy's global SFR while removing its gas (Figure \ref{fig:gas_and_sfr_global}), and the wind-enhanced star formation favors central disk regions (Figures \ref{fig:sf_radius_and_height} and \ref{fig:mass_flow_mdot_series}).
In this section, we evaluate the spatially-resolved SFR-mass relations --- a direct clue to the star formation microphysics (e.g., \citealt{kennicutt_star_2012}). 
We compare the relations between the RPS and isolated cases to characterize the physical conditions of ram pressure-enhanced star formation.

\subsection{Spatial Division Methodology and Radial Profiles}\label{subsec:radial_distribution}
When spatially resolving galactic regions, the sampling scales need to exceed certain minima for galactic star formation-mass scaling relations to hold \citep{kruijssen_uncertainty_2014,kruijssen_uncertainty_2018}; the selected sampling scales need to account for the incomplete statistical sampling of independent star-forming regions and the spatial drift between gas and stars. Empirically, this minimum spatial scale $\Delta x$ works out to be $\sim 1$ kpc for typical star-forming galaxy disks \citep[see][Fig. 2]{kruijssen_uncertainty_2014}. Here, we select the sampling scale to be 1 $\rm kpc^{2}$ to satisfy the validity of the scaling relations; and to match with typical scales ($0.75 - 1.2 ~ \rm kpc^{2}$) in high-angular-resolution observations in the local universe \citep{bigiel_star_2008,vulcani_gasp_2019,vulcani_gasp_2020,jimenez-donaire_vertico_2023}. 

For a given simulation snapshot, we divide the satellite disk into 1 $\rm kpc^{2}$-resolved patches, integrate the patches along the disk-height direction (for disk height $|z| \leq 2$ kpc), and calculate the projected SFR, gas, and stellar surface densities ($\Sigma_{\rm SFR}$, $\Sigma_{\rm gas}$, and $\Sigma_{*}$) of each patch. We focus on the galaxy group and cluster pericenter time frames, when RPS most effectively enhances/quenches the satellite star formation (Figure \ref{fig:gas_and_sfr_global}), and compare the wind cases with the isolated galaxy case (iso) at the corresponding times. Using 100 Myr windows (10 simulation outputs) produces larger samples of 1 kpc$^{2}$ regions with recent star formation (10 Myr) across the disk. Changing the number of outputs does not qualitatively affect our results. 
 Global properties of the selected patches are summarized in Table \ref{table:resolved_SFR_mass}. Throughout the Section, we will focus on the four pericentric cases in Table \ref{table:resolved_SFR_mass}: 13W, 14W, iso group, and iso cluster control, while iso pre-starvation is a special reference case for the star formation law comparison in \S \ref{subsec:local_KS} below.

\begin{deluxetable}{ccccc}\label{table:resolved_SFR_mass}
\tablecaption{Summary of the 1 $\rm kpc^{2}$ patch selection} 
\tablehead{\colhead{Case}  & \colhead{$t_{\rm peri}$} & \colhead{$\rm N_{patch,SF}$}    &   \colhead{$\rm SFR_{global}$} & \colhead{$\tau_{\rm dep}$} \\
\colhead{}  & \colhead{(Myr)} & \colhead{}  & \colhead{($M_{\odot}/\rm yr$)}   &    \colhead{(Gyr)}}
\decimalcolnumbers
\startdata
13W pericenter      &   2530-2630       & 1427 &   0.85    &  2.28  \\
iso group control   &   2530-2630       & 3016 &   0.38    &  7.12  \\
14W pericenter      &   2830-2930       &  447 &   0.21    &  1.70  \\
iso cluster control &   2830-2930       & 2933 &   0.32    &  8.16  \\
iso pre-starvation  &   580-680$^{(a)}$ & 2292 &   1.08    &  3.11  \\
\enddata
\tablecomments{(1) Simulation cases. (2) Simulation time periods that correspond to 13W and 14W pericenter passages. $^{(a)}$ The iso ``pre-starvation" time frame is selected to match the central plane $\rm M_{gas}$ of 13W pericenter (Figure \ref{fig:mass_flow_mdot_series}), see \S \ref{subsec:radial_distribution}. (3) The number of star-forming patches, where star formation is defined to have $\Sigma_{\rm SFR} > 10^{-6}~M_{\odot}/(\rm yr \cdot kpc^{2})$ per patch. (4) The total SFR of the star-forming patches averaged over the selected 100 Myr time period. (5) The gas depletion time defined as $\tau_{\rm dep} = M_{\rm gas}/\rm SFR$, where $M_{\rm gas}$ is the total gas mass in the patches, and SFR as in column 4.}
\end{deluxetable}

\begin{figure}
    \centering
    \includegraphics[width=0.95\linewidth]{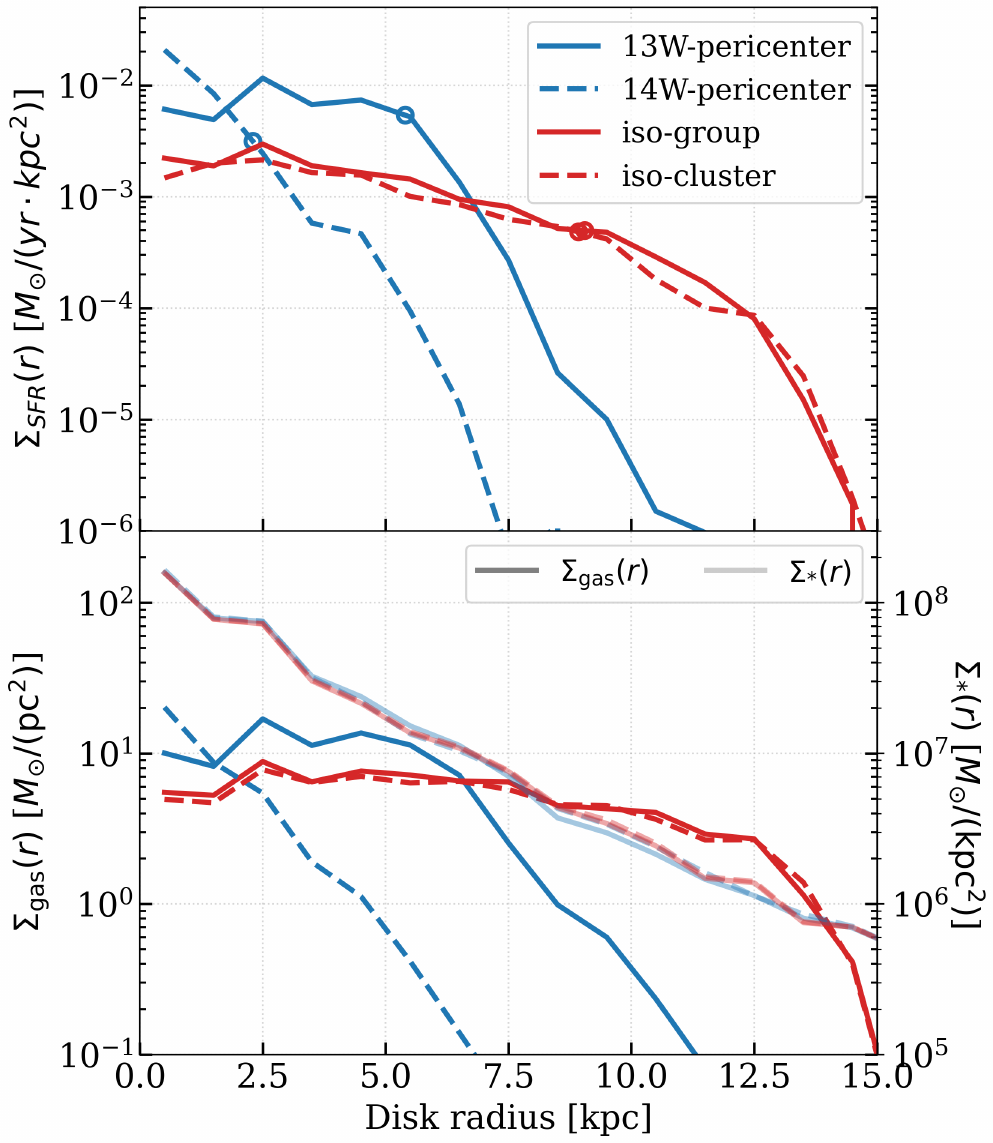}
    \caption{The SFR, gas, and stellar surface densities ($\Sigma_{\rm SFR}$, $\Sigma_{\rm gas}$, and $\Sigma_{*}$) radial profiles of the resolved 1 $\rm kpc^{2}$ patches. The four cases (Table \ref{table:resolved_SFR_mass}) are 13W and 14W at pericenters (solid and dashed in blue) and iso group and cluster control cases (same line styles in red). On the top panel, the open circles show the radii enclosing 95$\%$ of the $\Sigma_{\rm SFR}$. On the bottom panel, $\Sigma_{\rm gas}$ (deeper colors; left-hand y-axis) and $\Sigma_{*}$ (lighter colors; right-hand y-axis) are shown under the same scale, following the conventional units of each quantity.}
    \label{fig:radial_profile}
\end{figure}

The radial profiles of the resulting patches are shown in Figure \ref{fig:radial_profile}. The SFR profiles (top panel) closely resemble the respective gas profiles (bottom panel) in all cases. In the central few kpc, the wind cases (blue) are enhanced in both the SFR and gas densities relative to iso (red); at larger radii, the wind radial profiles show a steeper decrease with radius than iso. 
This is expected given our previous finding that ram pressure removes gas in the outer disk while driving gas into the central disk (Figure \ref{fig:mass_flow_mdot_series}). The 95$\%$ enclosing radii of the SFR, denoted by the open circles, show that star formation is more centrally-concentrated with increasing ram pressure, with the iso cases forming stars within $\sim$9.0 kpc (no ram pressure), 13W within $\sim$5.4 kpc (moderate ram pressure, enhanced SFR), and 14W within $\sim$2.3 kpc (strong ram pressure, approaching complete stripping). The time evolution in iso from the group to cluster pericenter times ($\sim$300 Myr duration; red solid and dashed curves) is due to star formation ``starvation", which has a relatively small impact on the radial profiles (and reduces the global SFR by $\sim 16\%$; see Table \ref{table:resolved_SFR_mass}).

The bottom panel of Figure \ref{fig:radial_profile} directly compares the gas and stellar surface density ($\Sigma_{\rm gas}$ and $\Sigma_{*}$) profiles. The y-axes are under the same physical scale following the conventional units of each quantity, as will be used in Sections \ref{subsec:local_KS} and \ref{subsec:local_sfr_star} and figures therein. In all cases, $\Sigma_{*}$ profiles (lighter lines) are greater than $\Sigma_{\rm gas}$ (deeper lines) within the inner $\sim$8 kpc region that encloses the majority of star formation. Unlike $\Sigma_{\rm gas}$ (or $\Sigma_{\rm SFR}$) that clearly distinguishes wind and iso, the $\Sigma_{*}$ profiles are consistent among all cases. This is because (i) ram pressure only directly impacts the gas disk and not the stellar disk; (ii) the formed stellar mass is low compared with the static stellar potential ($\Delta M_{\rm formed~star}/M_{\rm static~disk} < 3\%$; Figure \ref{fig:gas_and_sfr_global} and \S \ref{sec:methods}), therefore the static potential is dominated by the total $\Sigma_{*}$ (and dark matter) at all radii.

\subsection{SFR$-M_{\rm gas}$: The Kennicutt-Schmidt Relation}\label{subsec:local_KS}

We investigate the SFR$-M_{\rm gas}$ relation, also known as the Kennicutt-Schmidt (KS) relation \citep{schmidt_rate_1959,kennicutt_star_1989}, for the resolved 1 $\rm kpc^{2}$ patches. The KS relation is an empirical power-law between the observed SFR and gas surface densities, $\Sigma_{\rm SFR} = A \cdot \Sigma_{\rm gas}^{N}$. Physically, it is a proxy for how efficiently gas forms stars at given surface densities. In our suite of simulations, both the stripping and the isolated galaxy cases follow the same numerical star formation and feedback recipe (\citealt{goldbaum_mass_2015,goldbaum_mass_2016}; see \S \ref{sec:methods}), and differences on the $\Sigma_{\rm SFR} - \Sigma_{\rm gas}$ (KS) phase plane will directly reflect the impact of RPS.

\begin{figure*}
    \centering
    \includegraphics[width=1.0\linewidth]{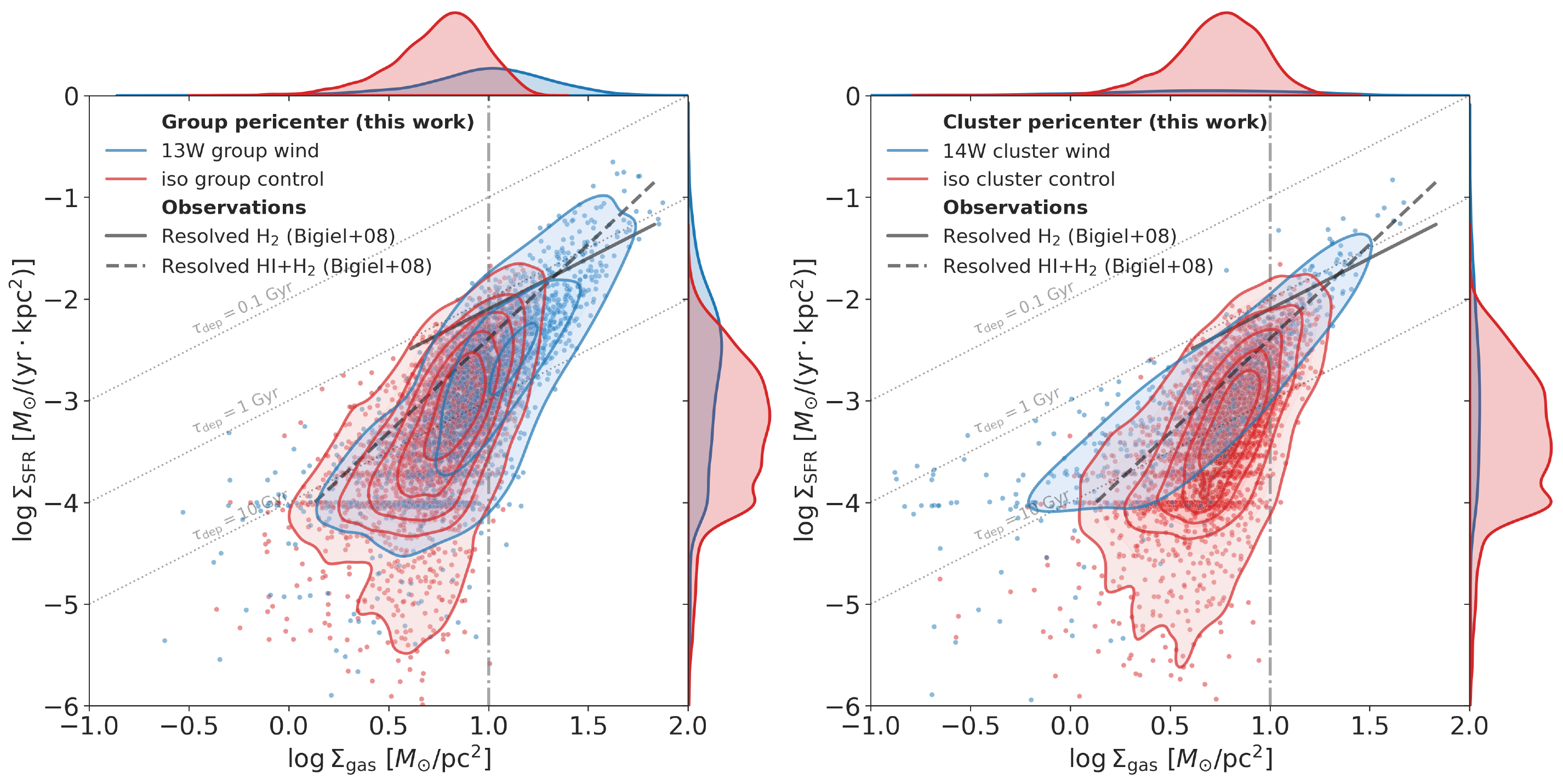}
    \caption{Spatially resolved SFR$-M_{\rm gas}$ ($\Sigma_{\rm SFR} - \Sigma_{\rm gas}$; ``Kennicutt-Schmidt") relation within disk height $|z| \leq 2$ kpc. The left and right panels show the 100 Myr duration of the galaxy group and cluster pericentric passages (or the corresponding times in iso; Table \ref{table:resolved_SFR_mass}), respectively. Each panel shows the $\Sigma_{\rm SFR} - \Sigma_{\rm gas}$ bivariate and one-dimensional (1D) distributions of 1 $\rm kpc^{2}$ patches in the iso (red) and wind (blue) simulations. Solid and dashed black lines show the spatially resolved KS power-law from \citet{bigiel_star_2008} over the observed $\Sigma_{\rm gas}$ ranges for H$_{2}$ and \HI+H$_{2}$ combined, 
    see \S \ref{subsec:local_KS} for details. Constant depletion time contours $t_{\rm dep}=0.1$, 1, 10 Gyr are annotated in gray dotted lines. Gray vertical line indicates the atomic-to-molecular gas density transition at $\Sigma_{\rm gas} \approx 10$ $M_{\odot}/\rm pc^{2}$ \citep{krumholz_atomic--molecular_2009}.}
    \label{fig:local_KS_law}
\end{figure*}

Figure \ref{fig:local_KS_law} shows the KS relation in the RPS and isolated galaxy disks at the group and cluster pericenters (Table \ref{table:resolved_SFR_mass}). For all cases, the gas densities are tightly correlated with SFR on the resolved scale, but the RPS cases populate a distinct phase space of high-density, high-SFR gas that is absent in iso. To quantify this excess, we first identify the 99.85 percentile surface density thresholds (empirically 3$\sigma$ upper limits) in iso using the 1D histograms of $\Sigma_{\rm SFR}$ and $\Sigma_{\rm gas}$. For each of these distributions, these upper limits are nearly identical at both the group and cluster pericentric times: 
$\log \Sigma_{\rm gas}/(M_{\odot}~\rm pc^{-2}) \gtrsim 1.2$ 
and $\log \Sigma_{\rm SFR}/(M_{\odot}~\rm yr^{-1}~kpc^{-2}) \gtrsim -1.8$. 
Many patches in 13W and 14W occupy the KS phase space beyond these upper limits in iso (blue scatter points; upper right corner of Figure 9); these are the dense gas excess in the RPS cases and have a significant contribution to the total SFR ($\sim$58\% in both 13W and 14W). 

For 13W, star formation from this dense gas (58\% or 0.5 $M_{\odot}/\rm yr$) is comparable with its SFR enhancement relative to iso ($\Delta$SFR$_{\rm group} \approx 0.45$ $M_{\odot}/\rm yr$; Table \ref{table:resolved_SFR_mass}). 
For 14W, ram pressure is strong enough to remove most of the surviving ISM and leads, ultimately, to a quenching of star formation. Despite 14W's dense gas excess, its number of star-forming patches ($N_{\rm patch,SF}$; Table \ref{table:resolved_SFR_mass}) has decreased to $\sim$30\% of 13W and $\sim$15\% of the iso control, resulting in its lowest total SFR of all cases. 
We will further discuss why gas and SFR surface densities are enhanced in the RPS cases in \S \ref{subsec:discussion_RP_SF}.

We showed in Figure \ref{fig:local_KS_law} that the RPS and iso cases populate different $\Sigma_{\rm gas}$ ranges. At the selected pericentric time frames, $\Sigma_{\rm gas}$ in iso primarily belongs in the \HI-dominated regime (the left of the dashed vertical line in both panels), while in 13W and 14W, it populates both \HIspace and H$_{2}$ regimes \citep{krumholz_atomic--molecular_2009}. The relatively low surface densities in iso are a direct result of the gradual gas consumption due to star formation and the feedback-driven outflows, also known as ``starvation" \citep{larson_evolution_1980,van_den_bosch_importance_2008,trussler_both_2020}. To evaluate the effect of RPS on the KS relation over a similar $\Sigma_{\rm gas}$ range, we identified an earlier time frame in iso 
(iso pre-starvation; see Table \ref{table:resolved_SFR_mass}), where the central disk gas mass --- and hence the highest $\Sigma_{\rm gas}$ --- is comparable to 13W at the pericenter (Figure \ref{fig:mass_flow_mdot_series}). This comparison will determine if the star formation efficiency at given gas densities is modified in the RPS cases.

\begin{figure}
    \centering
    \includegraphics[width=0.95\linewidth]{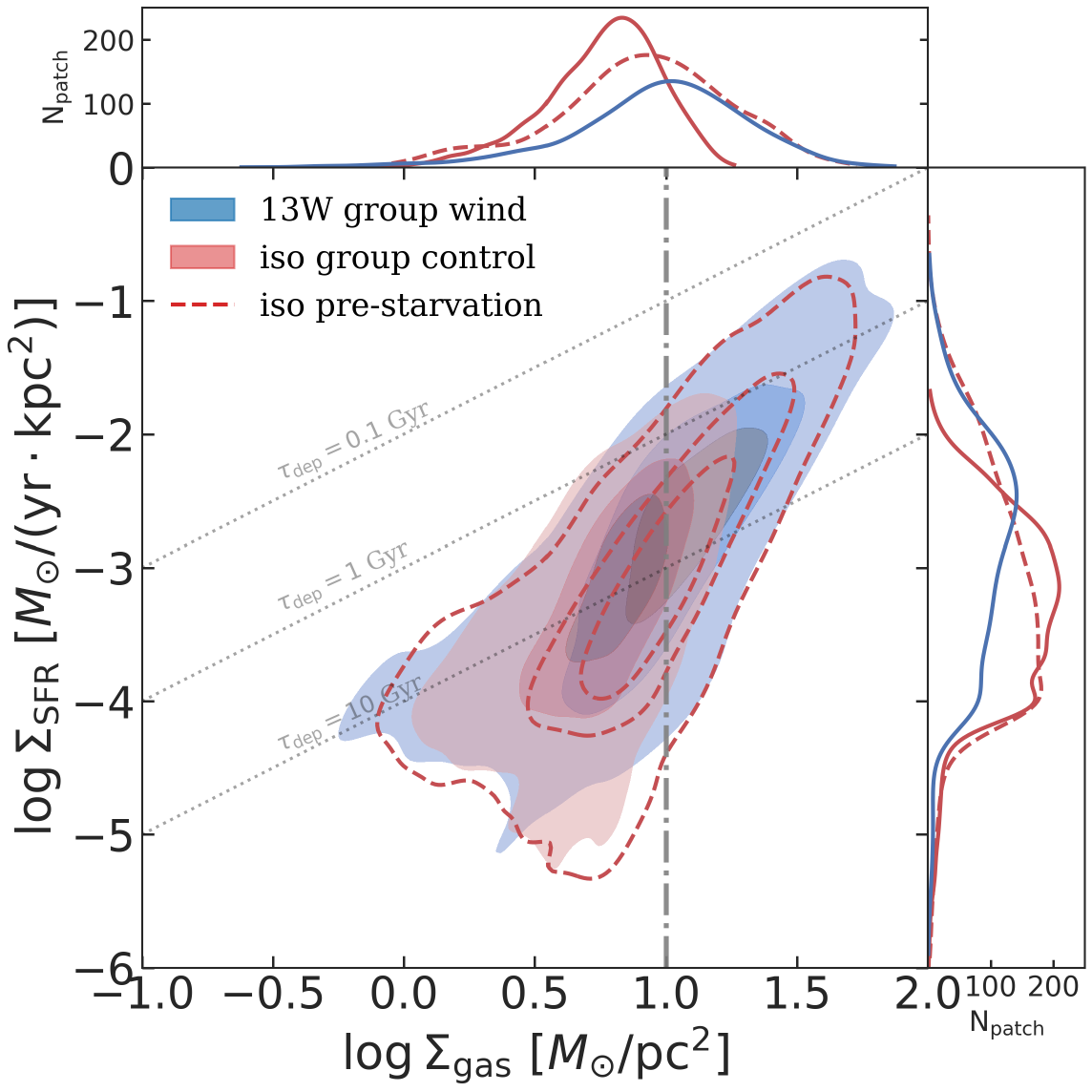}
    \caption{The effects of RPS versus starvation on the KS plane. The red and blue filled contours and solid 1D distribution lines are iso group control and 13W pericenter, respectively (as in Figure \ref{fig:local_KS_law}), and the red dashed lines are for iso at an earlier time (prior to $\sim$2 Gyr of starvation; see \S \ref{subsec:local_KS}). The constant depletion time contours are shown as in Figure \ref{fig:local_KS_law}.}
    \label{fig:KS_same_slope_transition}
\end{figure}

Figure \ref{fig:KS_same_slope_transition} shows the effect of RPS versus starvation on the KS plane by comparing three cases: 13W and iso at group pericenter time and iso pre-starvation. Starvation shifts the isolated galaxy to lower surface densities along the KS relation (red dashed versus red filled) via gas consumption and feedback-driven outflows. However, the iso pre-starvation case shares a very similar KS relation with the RPS case (red dashed versus blue filled), despite their distinct evolutionary history and gas morphology (Figures \ref{fig:gas_and_sfr_global} and \ref{fig:mass_flow_via_density_slice}). Judging from the similar KS relation, the star formation efficiency in RPS and iso cases remains the same at comparable $\Sigma_{\rm gas}$.

Independent of RPS, the resolved patches in our simulations show a KS power-law slope turnover from the \HIspace to H$_{2}$ regimes when sufficient dense gas exists as H$_{2}$ (Figures \ref{fig:local_KS_law} and \ref{fig:KS_same_slope_transition}). This is a direct reflection of our numerical star formation recipe \citep[see][Fig 5]{goldbaum_mass_2016}, which agrees with observational findings that the KS power-law slope transitions from superlinear in the atomic regime ($N_{\rm KS, \HI} > 1$ with poor correlation; \citealt{bigiel_star_2008,leroy_star_2008,kennicutt_star_2012}) to approximately linear in the molecular regime ($N_{\rm KS, H_{2}} \approx 1$; \citealt{krumholz_atomic--molecular_2009,heiderman_star_2010,krumholz_universal_2012,jimenez-donaire_vertico_2023}{; see the solid line in Figure \ref{fig:local_KS_law}). We also annotated the \HI+H$_{2}$ combined fitting result from \cite{bigiel_star_2008} in Figure \ref{fig:local_KS_law} (dashed line, N$_{\rm KS} \approx 1.8$); our simulations follow a mildly steeper slope in the atomic regime (N$_{\rm KS,\HI} \approx 2.0$), still well within the observational scatter \citep{bigiel_star_2008}. An exception to the overall consistent KS slope in our simulations is 14W at the lowest gas surface densities ($\Sigma_{\rm gas} \leq 0$ $M_{\odot}~\rm pc^{-2}$; see Figure \ref{fig:local_KS_law} right panel), which shows distinctively higher $\Sigma_{\rm SFR}$ than iso and hence a lower KS slope in the low-density \HIspace regime. The high $\Sigma_{\rm gas}$ (H$_{2}$-dominated) regime in 14W is similar to the other cases. We suspect that the low-density star-forming gas in 14W is driven by fast gas removal from RPS in recently star-forming regions.

\subsection{SFR$-M_{*}$: Strong Stripping and Disk Truncation}\label{subsec:local_sfr_star}
In this section, we investigate the SFR$-M_{*}$ relation, also known as the star formation main sequence relation (e.g., \citealt{schiminovich_uv-optical_2007,sargent_regularity_2014,speagle_highly_2014}), on the spatially-resolved $\Sigma_{\rm SFR}-\Sigma_{*}$ plane. We examine the impact of RPS using our simulations and make comparisons with \cite{vulcani_gasp_2020} (from the GAs Stripping Phenomena in galaxies ``GASP" survey; \citealt{poggianti_gasp_2017}). The SFR surface densities of the resolved simulation patches are identical to those in the KS relation (\S \ref{subsec:local_KS}), while the stellar surface densities are a combination of the static stellar disk and the formed star particles as outlined in \S \ref{subsec:radial_distribution}, which turns out to be highly consistent among all simulations (Figure \ref{fig:radial_profile}). The \cite{vulcani_gasp_2020} sample contains $\sim$1 $\rm kpc^{2}$ resolved patches within 30 RPS galaxies under various stripping stages in nearby clusters, along with 10 isolated control case galaxies of similar masses (see \citealt{vulcani_gasp_2019}). Our simulated galaxy with $M_{*} \approx 10^{9.8}$ $M_{\odot}$ lies well within the GASP sample range and is directly comparable.

\begin{figure*}
    \centering\includegraphics[width=1.0\linewidth]{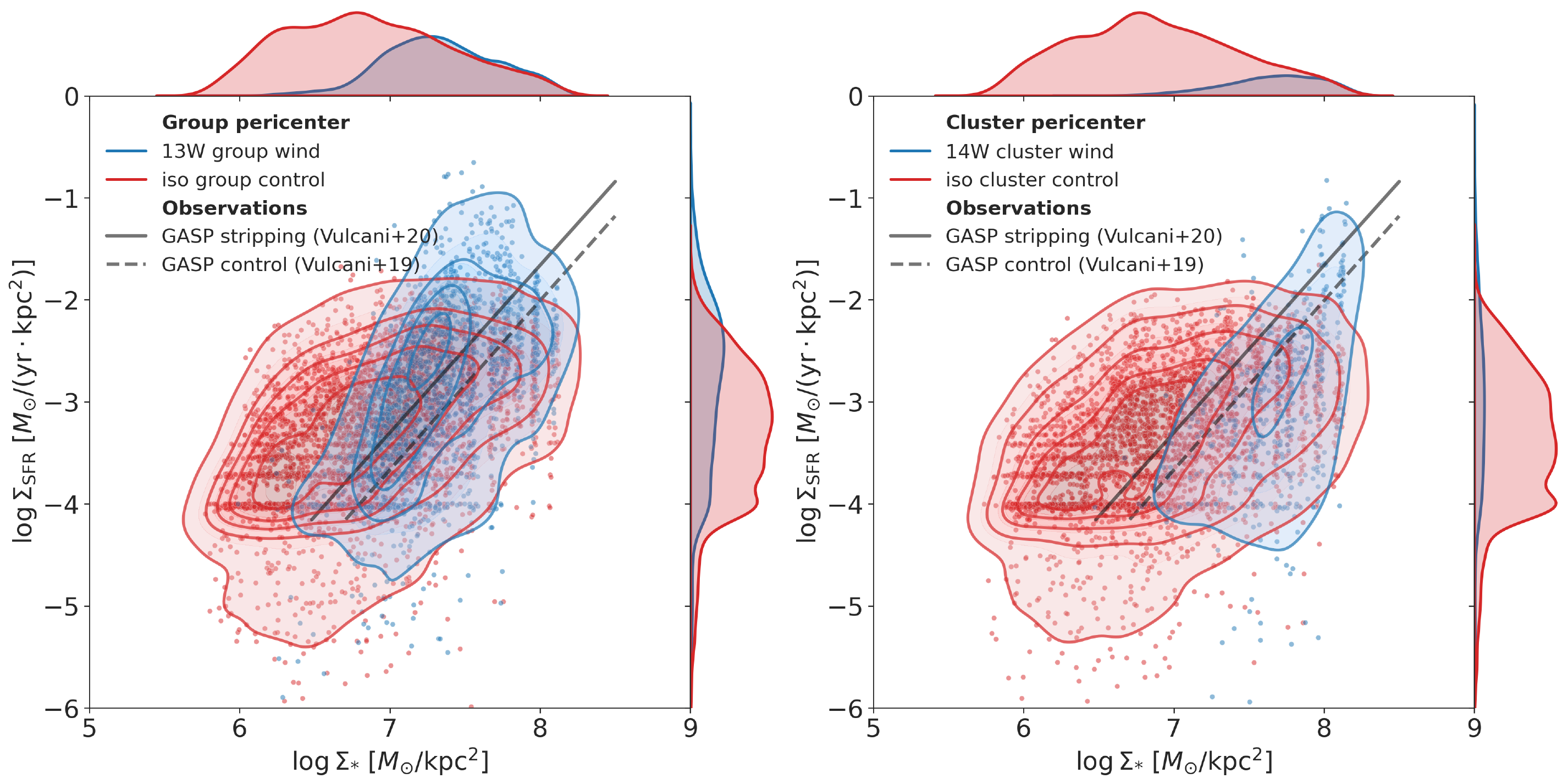}
    \caption{Spatially resolved SFR$-M_{*}$ ($\Sigma_{\rm SFR} - \Sigma_{*}$) relation within disk height $|z| \leq 2$ kpc, in the same style as Figure \ref{fig:local_KS_law}. The stellar surface densities are a combination of the static Plummer-Kuzmin potential (\S \ref{subsec:satellite_galaxy} and Table \ref{table:satellite_galaxy}) and the formed stars (as active particles) in the simulations. The black lines show the best-fit relations for the GASP stripping (solid; \citealt{vulcani_gasp_2020}) and isolated control (dashed; \citealt{vulcani_gasp_2019}) samples; both relations are shown over the observed $\Sigma_{\rm SFR}$, $\Sigma_{*}$ ranges, see \cite{vulcani_gasp_2020}.}
    \label{fig:local_sfr_star}
\end{figure*}

Figure \ref{fig:local_sfr_star} shows the $\Sigma_{\rm SFR} - \Sigma_{*}$ relation for the resolved patches, comparing 13W, 14W at their pericenters with the respective iso control cases in the same style as Figure \ref{fig:local_KS_law}. Since $\Sigma_{*}$ is a tight, monotonic function of disk radius (Figure \ref{fig:radial_profile}), it is an indicator for star formation location on the $\Sigma_{\rm SFR} - \Sigma_{*}$ phase plane. Two major effects of RPS can be identified from the differences between the wind and iso cases, (i) the truncation of the star-forming disk, shown by the high $\Sigma_{*}$ cutoffs for star formation, $\Sigma_{*} \approx 10^{6.5} (M_{\odot}~\rm kpc^{-2})$ in 13W and $10^{7} (M_{\odot}~\rm kpc^{-2})$ in 14W; (ii) the enhancement of star formation in the central disk regions, shown by the $\Sigma_{\rm SFR}$ excess at $\Sigma_{*} \gtrsim 10^{7.1} (M_{\odot}~\rm kpc^{-2})$ in 13W and $\Sigma_{*} \gtrsim 10^{7.6} (M_{\odot}~\rm kpc^{-2})$ in 14W. The disk truncation is more evident in 14W, where the ram pressure is higher, which is consistent with the radial profiles (Figure \ref{fig:radial_profile}).

In Figure \ref{fig:local_sfr_star}, we annotated the GASP best-fit power-law lines for the stripping (solid) and isolated control (dashed) samples \citep{vulcani_gasp_2020}, where the slopes are almost identical, but the stripping sample is $\sim$0.35 dex higher in $\Sigma_{\rm SFR}$ at all $\Sigma_{*}$. Our result is consistent with \cite{vulcani_gasp_2020} at high $\Sigma_{*}$ (central disk regions) that the spatially-resolved SFR is enhanced in the stripping cases. But we do not see a similar SFR enhancement at low $\Sigma_{*}$; instead, the low $\Sigma_{*}$ phase space is poorly populated in our wind cases due to disk truncation. We note that, however, the \cite{vulcani_gasp_2020} sample contains an ensemble of galaxies with a range of $M_{*}$, inclinations, and environments, while our simulations focus on one galaxy under different ram pressure strengths versus in isolation. The disk truncation and SFR enhancement in our wind cases are consistent with the ``Jstage=3" (strongest stripping) galaxies in \cite{vulcani_gasp_2020}, which do have a steeper fitted $\Sigma_{\rm SFR}$-$\Sigma_{*}$ slope. 

We also examined earlier pre-pericenter time frames in our simulations: qualitatively, the wind cases at earlier times (weaker stripping) show a similar truncation at lowest $\Sigma_{*}$ (disk edge) and a mild SFR enhancement at relatively high $\Sigma_{*}$ as those in Figure \ref{fig:local_sfr_star}. The transition $\Sigma_{*}$, where $\Sigma_{\rm SFR}$ in the wind cases becomes higher than in iso, increases with ram pressure strength, as expected for outside-in stripping. We will present the ``time-stacked" results from various RPS stages in \S \ref{subsec:obsn_predict}.

\section{Discussion}\label{sec:discussion}

\subsection{Impacts of RPS on star formation}\label{subsec:discussion_RP_SF}
Our key results in Sections \ref{sec:global_results} and \ref{sec:resolved_SF_mass} are,
\begin{enumerate}[nolistsep]
    \item In certain orbits, RPS can lead to an enhanced global star formation rate (SFR) in relatively gas-deficient galaxies.
    \item The SFR enhancement is driven by an excess of dense gas in the disk central regions, while the star formation efficiency at given gas surface densities (the Kennicutt-Schmidt relation) remains the same.
\end{enumerate}

There are two possible channels through which the SFR is enhanced in the galaxies undergoing ram pressure stripping: compression and mass transport, which are usually not separable \citep{tonnesen_star_2012,roediger_star_2014,troncoso-iribarren_better_2020,vulcani_gasp_2020}. The ISM at the ram pressure interface can be locally compressed, leading to a higher star formation rate and efficiency; global mass flows driven by ram pressure can redistribute gas in the disk and cause SFR enhancement. 

Our results show that RPS-driven mass transport, including fallback and radial inflows (\S \ref{subsec:gas_kinematics}; Figures \ref{fig:mass_flow_mdot_series} and \ref{fig:cumulative_central_flows}), is directly responsible for the central disk gas mass enhancement relative to iso during the $\sim$Gyr early stripping stage (Figure \ref{fig:mass_flow_mdot_series} upper right panel). The centralized ISM mass distribution in the RPS galaxies results in enhanced central surface densities ($\Sigma_{\rm gas}$ and $\Sigma_{\rm SFR}$; Figures \ref{fig:radial_profile} and \ref{fig:local_KS_law}), which account for the global enhancement of SFR in the stripping cases relative to iso (with only starvation). The signal of relative SFR enhancement exists for longer than Gyr timescales unless ram pressure becomes sufficient to remove the entire gas disk and quench the star formation (Figure \ref{fig:gas_and_sfr_global}).

The role of compression is more challenging to quantify and is often inferred indirectly. \cite{roediger_star_2014} found that compression, indicated by shock passages, can drive a local, short-lived SFR burst ($\sim$15 Myr), but it only impacts the low-density outer disk and has only a mild effect on the global SFR. \cite{choi_ram_2022} modeled a local patch of star-forming galactic disks under RPS and found a similar short-lived enhancement ($\sim$20 Myr; see their fig 13(c)) in the dense gas surface density ($\Sigma_{\rm gas,n_{H}>10~cm^{-3}}$), demonstrating vertical gas compression by the initial ram pressure passage. Compression would increase the gas volume density ($\rho_{\rm gas}$) without increasing the surface density integrated throughout the disk ($\Sigma_{\rm gas}$). If such compression happens, at comparable $\Sigma_{\rm gas}$, $\Sigma_{\rm SFR}$ will be systematically higher on the KS plane. However, our Figures \ref{fig:local_KS_law} and \ref{fig:KS_same_slope_transition} show that at the same gas surface densities, the stripping set follows the same KS relation as the iso set. This suggests that local compression is insufficient to account for the galaxy-scale SFR enhancement in our simulations.

Another search for compression in RPS galaxies in simulations was performed by \cite{troncoso-iribarren_better_2020}, which spatially divided satellite galaxy disks from the EAGLE simulations into leading and trailing halves (LH and TH) separated by the infall velocity vector. Under this LH-TH division, which maximizes the SFR asymmetry between the two halves of the satellite galaxies, \cite{troncoso-iribarren_better_2020} found that gas in the LH that tends to be more compressed, as inferred from higher average pressure, also has a higher star formation efficiency (defined as the total $\rm SFR/M_{\rm gas}$) compared with the TH. 
Here, we follow the methodology of \cite{troncoso-iribarren_better_2020} to further test for galaxy-scale effects of compression. We divide the galaxy disks based on a simple LH: $y<0$ and TH: $y \geq 0$ spatial criterion, given the infall velocity vector (Figure \ref{fig:mass_flow_via_density_slice}) and that the star-forming disk is thin throughout the simulations (Figure \ref{fig:sf_radius_and_height}). In the discussion hereafter, we will assume that the LH is under higher compression than the TH because of ram pressure. The disparity (or the lack of) in mass and star formation between the LH and TH will help disentangle the effects of mass transport and compression.

\begin{figure}
    \centering
    \includegraphics[width=0.95\linewidth]{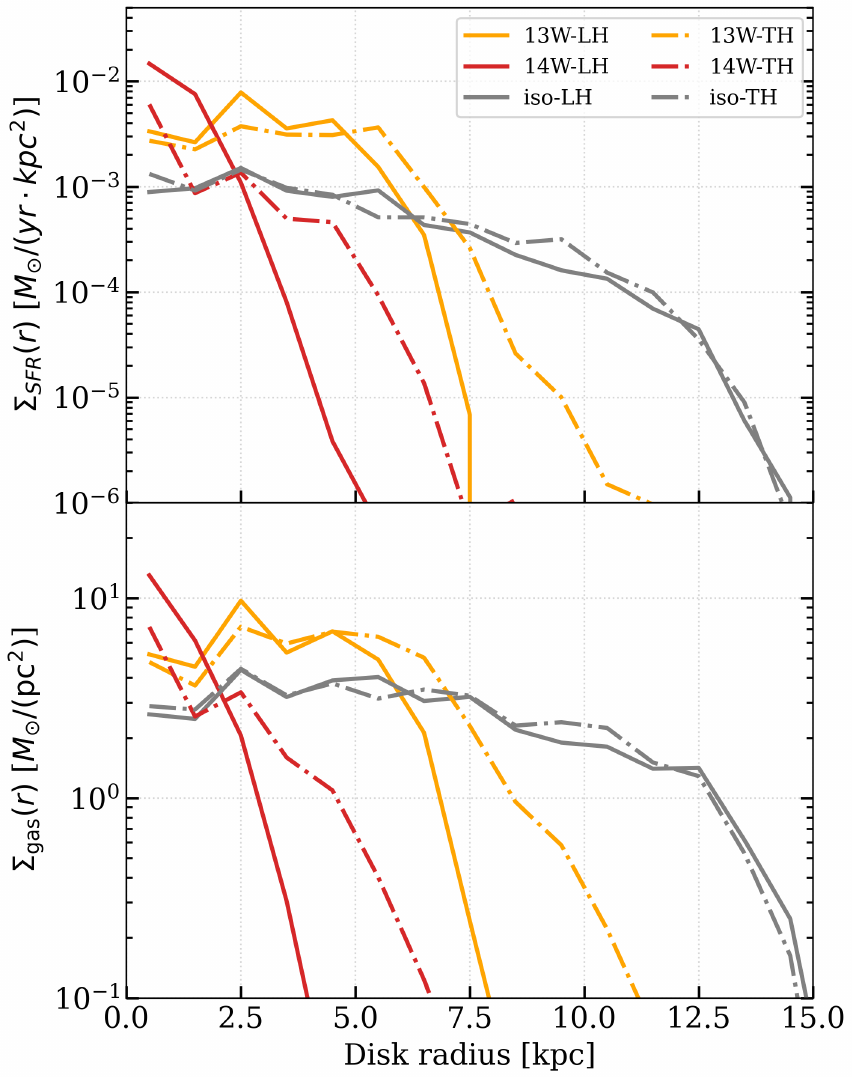}
    \caption{The SFR and gas surface density radial profiles (described in \S \ref{subsec:radial_distribution}) for the wind-LH and TH of the disks. The simulations are color-coded as in \S \ref{sec:global_results}, and we omitted 12W where the ram pressure has negligible impacts on the gas or SFR (Figure \ref{fig:gas_and_sfr_global}). For each wind simulation, we averaged over the 100 Myr closest to the pericenter (10 outputs), as in Table \ref{table:resolved_SFR_mass}; for iso, we selected the group pericenter time, which yields a largely consistent profile with the cluster pericenter time (Figure \ref{fig:radial_profile}).}
    \label{fig:radial_profile_LH_TH}
\end{figure}

Figure \ref{fig:radial_profile_LH_TH} shows the SFR and gas radial profiles of the wind and iso runs (similar to Figure \ref{fig:radial_profile}), distinguishing the LH (solid lines) and TH (dashed lines) of the disks. As expected, the surface density profiles for the isolated galaxy control case are an equal division between the two halves at all radii. In the wind runs, the surface densities in the LH consistently show a steeper decrease with disk radius than the TH; the disk radius at which the two halves diverge decreases with ram pressure. The more extended low $\Sigma_{\rm SFR}$, $\Sigma_{\rm gas}$ material in the TH is caused by the asymmetric disk morphology\footnote{Our $|z| \leq 2$ kpc disk height selection excludes tail contamination.} under an inclined wind (Figure \ref{fig:mass_flow_via_density_slice}). Within the central few kpcs of the disk, the surface density profiles show a mild LH excess in 13W and a stronger excess in 14W. However, we found that the 14W signal oscillates over time\footnote{Figure \ref{fig:time_series_LH_TH} shows the temporal oscillations of the LH-TH differences, which can explain the 14W central disk LH excess in Figure \ref{fig:radial_profile_LH_TH}. If we select, e.g., 100 Myr prior to the pericenter (t=2730-2830 Myr), the LH densities become lower than the TH in the central disk. But the mild LH excess in the 13W central disk is independent of the oscillations during the few hundred Myr approaching the pericenter because the accumulation of fallback gas mildly favors the LH (Figure \ref{fig:grav_potential_fallback_and_mixing}).}, which is likely due to the orbit of the few dense clouds remaining before complete stripping instead of compression.

\begin{figure}
    \centering
    \includegraphics[width=0.95\linewidth]{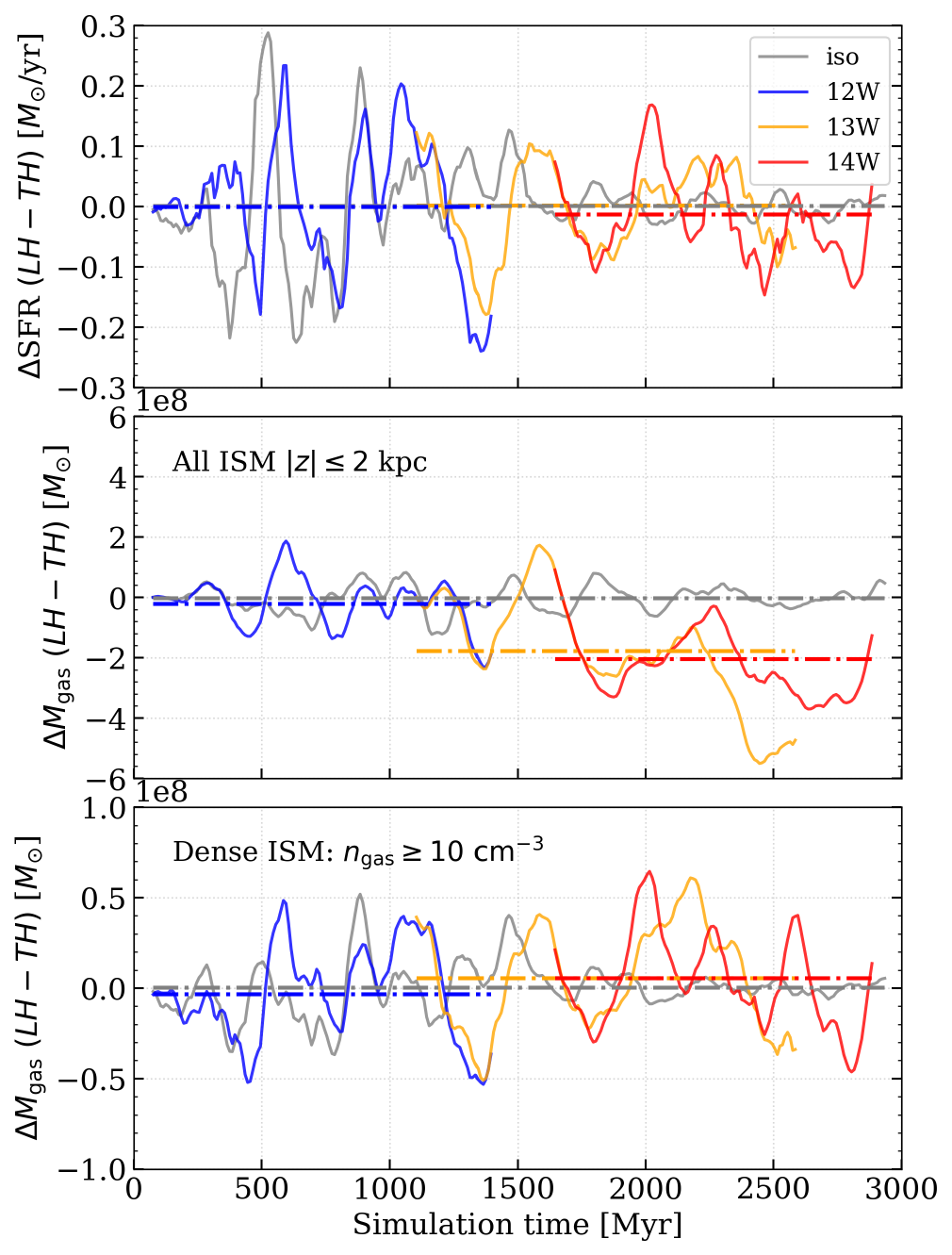}
    \caption{The SFR and gas mass differences between the LH and TH versus simulation time. The three panels from top to bottom show the differences (LH minus TH) in SFR, the disk ISM mass, and the dense ISM mass (where number density exceeds the star formation threshold; see \S \ref{sec:methods}), respectively. In each panel, the solid curves show the running means over $100$ Myr, and the horizontal dash-dotted lines show the time averages of individual simulations.}
    \label{fig:time_series_LH_TH}
\end{figure}

Figure \ref{fig:time_series_LH_TH} shows the time evolution of the LH-TH disparity. The time-dependent differences in the SFR and gas mass are subject to oscillations due to disk rotation and epicyclic motions (as previously shown in Figure \ref{fig:sf_radius_and_height}; also see \citealt{tonnesen_gas_2009}), which always average to $\sim$0 in the absence of ram pressure; see the time averages of iso (horizontal gray dash-dotted line). Under ram pressure, the time-averaged SFR remains close to equal between the two halves (top panel); the disk ISM mass shows a strong excess in the TH under intermediate and strong ram pressure (middle panel); but the dense, star-forming ISM (bottom panel) mass is again almost equal between the two halves. The temporal trends of the SFR generally follow those of the dense ISM; they are much less sensitive to the total ISM, which acquired the strongest LH-TH disparity from RPS.

The primary effect of RPS (with an edge-on component) is generating an excess of low-density gas in the TH that has a low contribution to the global SFR (Figure \ref{fig:radial_profile_LH_TH}). The SFR and dense ISM of the two halves, although subject to temporal oscillations, show close to equal time-averaged values and no trend with respect to ram pressure (Figure \ref{fig:time_series_LH_TH}). Our finding of the RPS-driven gas excess in the TH agrees with \cite{troncoso-iribarren_better_2020}, but our interpretation of this asymmetry differs. Since \cite{troncoso-iribarren_better_2020} defined the star formation efficiency of each half as the mass-weighted $\rm SFR/M_{\rm gas}$, the TH efficiency may be biased by the excess of non-star forming gas, appearing as an efficiency enhancement in the LH. We showed that the dense ISM responsible for star formation shows no such disparity (Figure \ref{fig:time_series_LH_TH}), indicating that the likely more compressed LH has the same efficiency as the TH.

To conclude, compression is not the direct cause of the ram pressure-induced SFR enhancement in our simulations, judging from two independent tests,} (i) the spatially resolved SFR surface densities ($\Sigma_{\rm SFR}$) in the stripping set show no systematic enhancement at comparable $\Sigma_{\rm gas}$ (inferred from KS relation; Figures \ref{fig:local_KS_law} and \ref{fig:KS_same_slope_transition}), (ii) galaxy-scale global properties, SFR and dense ISM mass, show no enhancement in the LH where compression is stronger. Instead, the RPS-induced mass flows (Figures \ref{fig:mass_flow_mdot_series} and \ref{fig:cumulative_central_flows}) account for the centralized mass and SFR profiles (enhanced central surface densities; Figures \ref{fig:radial_profile} and \ref{fig:radial_profile_LH_TH}), which supports mass transport as the direct mechanism for the SFR enhancement.


\subsection{Predictions for observations}\label{subsec:obsn_predict}

Here we predict RPS observables based on the simulation results, where Section \ref{subsubsec:gas_fraction_discuss} focuses on the surviving gas in the disk and Section \ref{subsubsec:sfr_mass_time_stack_discuss} on the local SFR-mass relations. We will discuss our predictions in the context of recent environmental surveys, GASP \citep{poggianti_gasp_2017,moretti_high_2020,vulcani_gasp_2020} and VERTICO \citep{brown_vertico_2021,jimenez-donaire_vertico_2023}.

\subsubsection{Surviving gas in the disk: the dense gas ratio and the gas mass fraction}\label{subsubsec:gas_fraction_discuss}

\begin{figure*}
    \centering
    \includegraphics[width=0.95\linewidth]{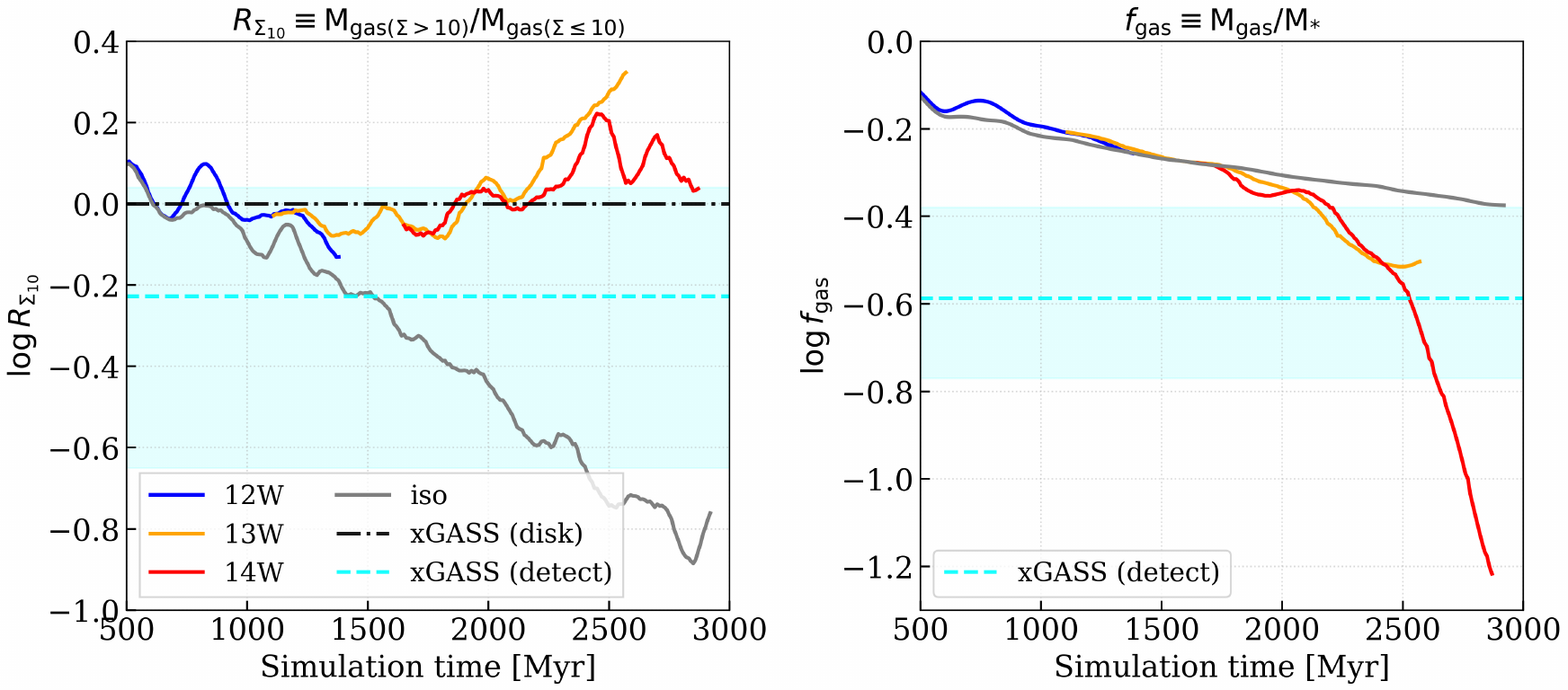}
    \caption{Time evolution of the dense gas ratio $R_{\Sigma_{10}}$ (left) and the gas mass fraction $f_{\rm gas}$ (right). For the dense gas ratio, we employed a simplified $\Sigma_{\rm gas}=10$ $M_{\odot}/\rm pc^{2}$ cut to distinguish the \HI- and H$_{2}$-dominated gas; see \S \ref{subsec:obsn_predict}. The dash-dotted line on the left panel shows the average $M_{\rm H_{2}}/M_{\rm H\,\textsc{i}}$ ratio for the disk regions of the xGASS sample  
    (see Fig 3 of \citealt{moretti_high_2020}). The cyan dashed lines and shadings on both panels show the median and the first to third quartile ranges of the xGASS sample \citep{saintonge_xcold_2017,catinella_xgass_2018} at a comparable stellar mass range ($9.6 < \log M_{*}/M_{\odot} < 9.9$). All simulation quantities are for the disk only (disk height $|z| \leq 2$ kpc; excluding the stripping tails).}
    \label{fig:gas_fraction_time_evol}
\end{figure*}

We describe the surviving gas using two global quantities: the dense gas ratio ($R_{\Sigma_{10}}$) and the gas mass fraction ($f_{\rm gas}$) within the simulated disk. The dense gas ratio is an estimate for the H$_{2}$ to \HIspace mass ratio, defined as $R_{\Sigma_{10}} \equiv M_{\rm gas (\Sigma_{gas} > 10)} / M_{\rm gas (\Sigma_{gas} \leq 10)}$, where $\Sigma_{\rm gas} = 10$ $M_{\odot}/ \rm pc^{2}$ is adopted as an empirical atomic-to-molecular transition density (\citealt{krumholz_atomic--molecular_2009,kennicutt_star_2012}; also see \S \ref{subsec:local_KS}). The ratio $R_{\Sigma_{10}}$ is not direct modeling of $M_{\rm H_{2}}/M_{\rm H\,\textsc{i}}$, but it self-consistently compares the molecular- and atomic-dominated gas masses in our simulations. The gas mass fraction is defined as $f_{\rm gas} \equiv M_{\rm gas}/M_{*}$, where $M_{\rm gas}$ is the total gas mass within the disk and $M_{*}$ includes the static stellar potential (\S \ref{sec:methods}) and the formed star particles. Figure \ref{fig:gas_fraction_time_evol} shows the time evolution of both quantities, for which we consistently selected disk height $|z| \leq 2$ kpc, hence excluding most of the unbound gas in the tail.

In Figure \ref{fig:gas_fraction_time_evol}, we annotated the observational results from the extended GALEX Arecibo SDSS Survey (xGASS; \citealt{catinella_xgass_2018}
) as a reference. The comparison sample we adopted (cyan on both panels, dashed line for median, shading for first to three quartiles) is a subset of 21 xGASS galaxies with comparable stellar masses ($M_{*} \in 10^{9.6-9.9} M_{\odot}$) to our simulated satellite galaxy, and with detections in both CO and \HIspace \citep{saintonge_xcold_2017,catinella_xgass_2018}. We consistently adopted $M_{\rm H_{2}}/M_{\rm H\,\textsc{i}}$ as the dense gas ratio (left panel) for the observational samples. Additionally, we showed the average dense gas ratio of xGASS disk regions (dashed-dotted line; following \citealt{moretti_high_2020}), which has an additional disk radius selection \citep{wang_xgass_2020} that results in a higher dense gas ratio.

In our simulations, the dense gas ratio $R_{\Sigma_{10}}$ in the RPS cases is consistently higher than that in the isolated galaxy case (Figure \ref{fig:gas_fraction_time_evol} left panel). The ratio $R_{\Sigma_{10}}$ increases with time in 13W and 14W, as opposed to the clearly decreasing trend in iso. The decreasing trend in iso is a result of starvation (Figure \ref{fig:KS_same_slope_transition}): gas depletion due to star formation and feedback favors the high $\Sigma_{\rm gas}$ regions with high local $\Sigma_{\rm SFR}$, reducing the ratio of the denser (e.g., $\Sigma_{\rm gas} > 10$ $M_{\odot}/\rm pc^{2}$; H$_{2}$-dominated) gas. While in the RPS cases, gas removal favors the low $\Sigma_{\rm gas}$ regions where the gravitational restoring force is weakest (``outside-in stripping"), and the disk central regions can be replenished by the ram pressure-driven mass flows like fallback and radial inflows (\S \ref{subsec:discussion_RP_SF}) --- both mechanisms increasing the dense gas ratio within the disk. Compared at the pericenter times (Table \ref{table:resolved_SFR_mass}), $R_{\Sigma_{10}}$ in 13W and 14W are $1.04$ and $0.87$ dex higher than iso (a factor of $11.0$ and $7.5$), respectively. 

Our result that RPS can increase the dense gas ratio agrees with \cite{moretti_high_2020}, which found a factor of 4 to $\sim$100 higher $M_{\rm H_{2}}/M_{\rm H\,\textsc{i}}$ ratios for three GASP jellyfish galaxies (undergoing active RPS) than the xGASS disk control sample (dash-dotted line in Figure \ref{fig:gas_fraction_time_evol}). \cite{moretti_high_2020} suggested that a more efficient conversion of neutral into molecular gas in these jellyfish galaxies can explain their significantly higher molecular mass ratios. However, the $R_{\Sigma_{10}}$ trends in our simulations can be explained by the different gas depletion models under starvation versus RPS described above, independent of the \HI-H$_{2}$ conversion. Despite the different definitions, our $R_{\Sigma_{10}}$ values are comparable with the observational $M_{\rm H_{2}}/M_{\rm H\,\textsc{i}}$ values of the xGASS sample (Figure \ref{fig:gas_fraction_time_evol}); we do not see the high $M_{\rm H_{2}}/M_{\rm H\,\textsc{i}} \gtrsim 10$ ($\log R_{\Sigma_{10}} > 1$) of the three jellyfish galaxies in \cite{moretti_high_2020}. 

The evolution of the gas mass fraction $f_{\rm gas}$ (Figure \ref{fig:gas_fraction_time_evol} right panel) closely follows that of $M_{\rm gas}$ (Figure \ref{fig:gas_and_sfr_global}), because the stellar mass evolution is relatively minimal throughout the simulations ($M_{*}=10^{9.7-9.8}$ $M_{\odot}$). The initial condition of $\log f_{\rm gas} \approx 0$ we adopted (\S \ref{subsec:satellite_galaxy}) is $\sim$0.5 dex higher than the xGASS average value ($0.01 < z < 0.05$ galaxies, \citealt{catinella_xgass_2018}, but within 1$\sigma$ of \citealt{calette_hi-_2018}), otherwise the $f_{\rm gas}$ evolution throughout the iso simulation is within the observational scatter of xGASS (cyan shading in Figure \ref{fig:gas_fraction_time_evol}). We take $f_{\rm gas}(t)$ in iso as the reference gas fraction in our simulations: 12W is in overall agreement with iso, 13W at the group pericenter is mildly lower ($\Delta \log f_{\rm gas} \approx -0.16$ dex), and 14W at the cluster pericenter significantly lower ($-0.84$ dex). As expected, direct removal by RPS decreases the total $M_{\rm gas}$ and hence $f_{\rm gas}$ in the group and cluster cases. But in the group case, a $-0.16$ dex deviation from prediction is within the typical observational scatter ($0.2-0.3$ dex) of such relations (see Figure \ref{fig:gas_fraction_time_evol} and \citealt{cortese_dawes_2021} fig 2). The satellite at 13W pericenter has a reduced gas fraction but still belongs to the gas normal regime, while at 14W, it is gas deficient during the final $\sim$400 Myr approaching the pericenter.

For the simulation cases with observable gas stripping morphology (13W and 14W), RPS always reduces $f_{\rm gas}$ in the galaxy disks. This means that under RPS, despite the mass transfer channels that can potentially replenish the dense/central-disk gas, stripping of the low-density/outer-disk gas dominates the global mass evolution (see, e.g., Figure \ref{fig:mass_flow_mdot_series}). 
But when we account for the total gas in the disk \textit{and tail} (tail gas potentially unbound), we find, similarly to \cite{moretti_high_2020}, that $f_{\rm gas,disk+tail}$ is similar between the RPS and iso cases.

To summarize \S \ref{subsubsec:gas_fraction_discuss}, first, RPS with an edge-on component tends to increase the dense gas ratio in the disk, while starvation decreases it. This could explain the observed higher molecular-to-atomic gas ratio ($M_{\rm H_{2}}/M_{\rm H\,\textsc{i}}$) in jellyfish galaxies \citep{moretti_high_2020} without requiring a substantially higher \HI-H$_{2}$ conversion efficiency. Second, 
RPS (unsurprisingly) reduces the gas mass fractions in the disk, even where the global SFR is enhanced. Where ram pressure at the orbital pericenter is insufficient to remove the densest ISM (13W), the $f_{\rm gas}$ reduction can be mild, maintaining the stripped galaxy in the gas normal regime ($< \pm 0.3$ dex). The remaining gas in the disk at 13W pericenter will likely be perturbed by galaxy-galaxy gravitational interactions, which are expected to be effective in group environments; see \S \ref{subsec:limitations}.%

\subsubsection{RPS signatures on the local SFR-mass relations}\label{subsubsec:sfr_mass_time_stack_discuss}

High angular resolution observations have enabled the direct mapping of galactic star formation laws on small scales \citep{bigiel_star_2008,leroy_star_2008,kennicutt_star_2012}. Some recent programs include the PHANGS (Physics at High Angular resolution in Nearby GalaxieS) survey for nearby galaxies \citep{leroy_phangs-alma_2021,lee_phangs-hst_2022}, the VERTICO survey for Virgo cluster galaxies \citep{brown_vertico_2021,jimenez-donaire_vertico_2023}, and the GASP survey for environmentally selected jellyfish galaxies \citep{poggianti_gasp_2017,jaffe_gasp_2018,vulcani_gasp_2020,moretti_high_2020}. RPS is one of the main environmental processes in the environmentally-selected samples (e.g., GASP and VERTICO), but the assessment of the RPS impact often faces several challenges: (i) the inevitable mixture of sample stellar masses and inclination angles
, (ii) the difficulties of constraining the environment (e.g., ICM densities) and the satellite orbits
, and (iii) the complex gravitational effects that could coexist with RPS
. Here, we use our simulation suite, which focuses on a single galaxy across various environments undergoing RPS and no tidal effects, to make predictions for the observational ``RPS signatures" on the local SFR-mass relations.

\begin{figure*}
    \centering
    \includegraphics[width=0.9\linewidth]{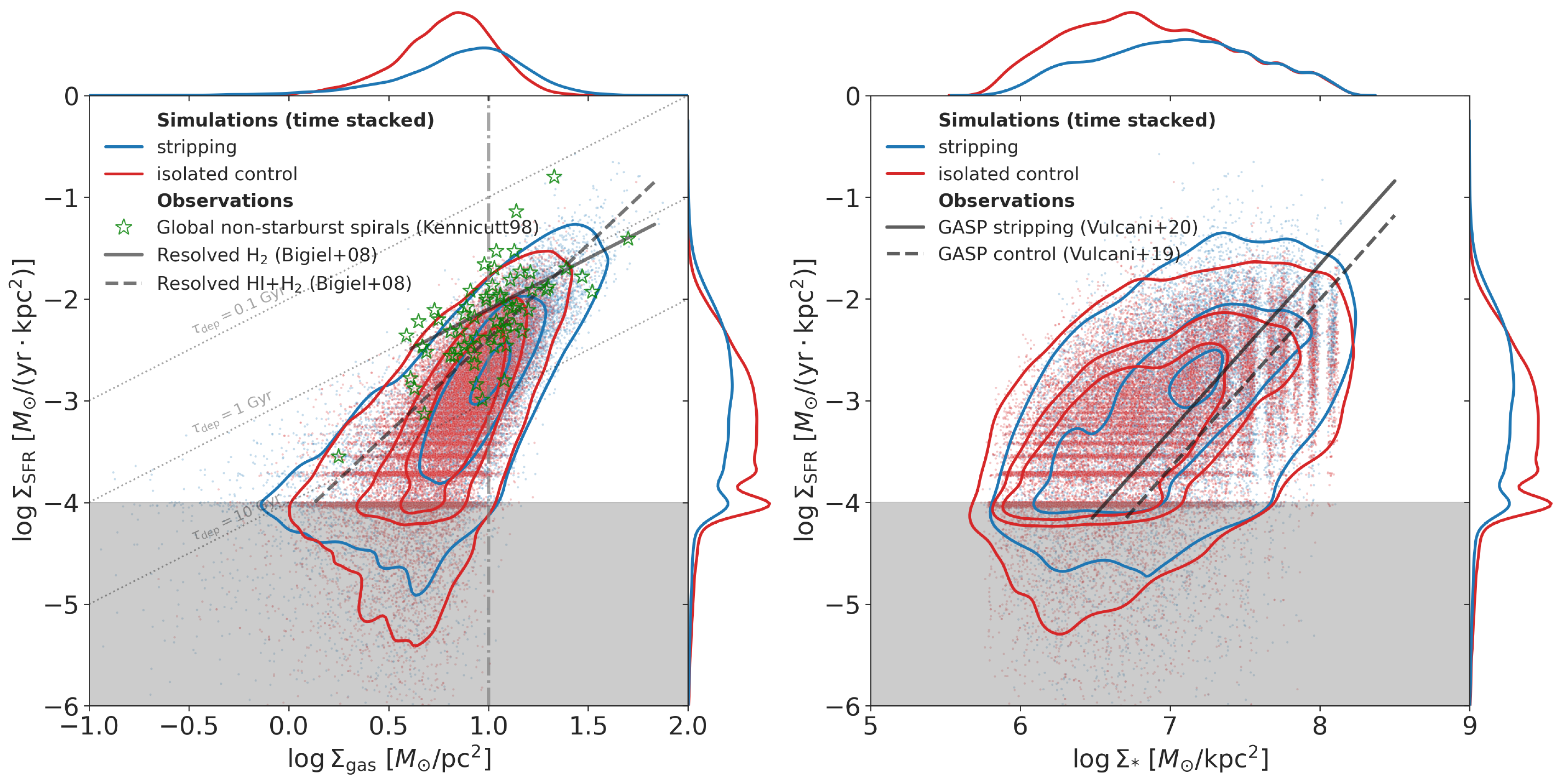}
    \caption{Spatially resolved SFR-mass relations (KS and star formation main sequence; see \S \ref{sec:resolved_SF_mass}), similar to Figures \ref{fig:local_KS_law} and \ref{fig:local_sfr_star}. Here for the observational predictions, we stacked 900 Myr of simulation data (90 outputs), covering more than 3 dex of ram pressure strengths in the wind runs (\S \ref{subsec:obsn_predict}). The shaded regions in both panels are where $\log \Sigma_{\rm SFR}$ is below current observational limits ($\sim$-4 dex). On the left panel, we additionally show the global KS relation of the non-starburst spiral galaxies from \citet{kennicutt_global_1998}, see \S \ref{subsubsec:sfr_mass_time_stack_discuss}.}
    \label{fig:sfr_mass_time_stacked}
\end{figure*}

To create mock observational datasets, we selected 900 Myr of simulation data (90 outputs) that cover more than 3 dex of $P_{\rm ram}$ in the wind runs (Figure \ref{fig:ram_pressure_evolution}), ranging over 1.5 Gyr in simulation time. Stacking the selected data creates two datasets, the stripping set and the isolated control set. This is equivalent to observing an ensemble of galaxies at $M_{*} \approx 10^{9.7-9.8}$ $M_{\odot}$ undergoing various stages of RPS (stripping set: Milky Way-like to cluster pericenter environments) or starvation (isolated control set: $\sim$1.5 Gyr duration). Figure \ref{fig:sfr_mass_time_stacked} shows the local SFR-mass relations for the two sets on 1 kpc$^{2}$ scale, following the methodology described in \S \ref{sec:resolved_SF_mass}. In Figure \ref{fig:sfr_mass_time_stacked}, we shaded the low SFR regions where $\log \Sigma_{\rm SFR}$ are below observational limits ($\sim$-4 dex, e.g., \citealt{leroy_estimating_2012,kennicutt_star_2012,vulcani_gasp_2020}). The over-densities in the simulation data at certain low $\Sigma_{\rm SFR}$ is a numerical effect due to our star particle mass resolution (e.g., the lowest horizontal over-density corresponds to the $\Sigma_{\rm SFR}$ from a single star particle of $\approx 800$ $M_{\odot}$). To guide observational comparison, we show the global KS relation for the 61 non-starburst spiral galaxies from \citet{kennicutt_global_1998} on the left panel in addition to the resolved KS relations from \citet{bigiel_star_2008}; also see Figure \ref{fig:local_KS_law}. The discontinuous sampling at high $\Sigma_{*}$ (right panel) is caused by the limited 1 kpc$^{2}$ spatial patch number in the disk center. 

The resolved KS relation for the time-stacked stripping and iso sets is remarkably consistent (Figure \ref{fig:sfr_mass_time_stacked} left panel). When observed at different snapshots in time (e.g., in Figures \ref{fig:local_KS_law} and \ref{fig:KS_same_slope_transition}), the galaxy populates different $\Sigma_{\rm gas}$ ranges, which is closely correlated with the local $\Sigma_{\rm SFR}$ (\S \ref{subsec:local_KS}). In the time-stacked view, where the $\Sigma_{\rm gas}$ ranges become similar between the two sets, the underlying KS relation is in overall agreement. Our finding is consistent with \cite{jimenez-donaire_vertico_2023} (VERTICO) results that the local KS relation agrees between an ensemble of Virgo RPS satellites and their isolated field counterparts, which suggests that RPS does not directly affect the local star formation efficiency within the gas.

The main difference between the stripping and isolated control sets is best seen in the 1D distributions of $\Sigma_{\rm gas}$ and $\Sigma_{\rm SFR}$. The stripping set reaches higher maximum surface densities and is truncated at low surface densities; the peaks in both $\Sigma_{\rm gas}$ and $\Sigma_{\rm SFR}$ distribution are at higher values compared with the isolated set. The difference can be explained by a combination of low-density gas removal and high-density gas replenishment in the stripping set; see the evolution of the dense gas ratio (\S \ref{subsubsec:gas_fraction_discuss}). Another relatively minor difference is that the stripping set reaches higher $\Sigma_{\rm SFR}$ at low $\Sigma_{\rm gas}$ ($<0$ dex), which, as also discussed above, is likely due to fast gas removal by RPS in still star-forming regions. The signals of high $\Sigma_{\rm SFR}$ at low $\Sigma_{\rm gas}$ only occur shortly before the complete removal of gas (cluster pericenter in Figure \ref{fig:local_KS_law}).

The time-stacked star formation main sequence relation (Figure \ref{fig:sfr_mass_time_stacked} right panel) shows a mild $\Sigma_{\rm SFR}$ enhancement over a range of $\Sigma_{*}$ in the stripping set. Independent of RPS or starvation, the $\Sigma_{*}$ radial profiles remain monotonic (Figure \ref{fig:radial_profile}), so the local $\Sigma_{*}$ can be used as an indicator of disk radii. Under increasing ram pressure, the $\Sigma_{*}$ threshold where the wind SFR exceeds iso SFR increases (corresponding disk radius decreases), so the time-stacked result here shows a smoother $\Sigma_{\rm SFR}$ enhancement that extends to lower $\Sigma_{*}$ (larger radii) compared with the group and cluster pericenter cases (Figure \ref{fig:local_sfr_star}). This agrees with the \cite{vulcani_gasp_2020} (GASP) finding that $\Sigma_{\rm SFR}$ can be enhanced over a range of $\Sigma_{*}$ when accounting for various stripping stages. However, at the lowest $\Sigma_{*}$ end, where \cite{vulcani_gasp_2020} found SFR enhancement in the stripping sample, our simulations always show disk truncation (lowest $\Sigma_{*}$ not populated in the stripping set; \S \ref{subsec:local_sfr_star}), instead of SFR enhancement. We note that since we focused on the disk region ($|z| \leq 2$ kpc), our sampled patches are free of star-forming clumps in the tail, which are shown to have a higher $\Sigma_{\rm SFR}$ than the disk at low $\Sigma_{*}$ \citep{vulcani_gasp_2020}.

Our predictions in \S \ref{subsubsec:sfr_mass_time_stack_discuss} can be summarized as follows. When observing a large ensemble of RPS and isolated galaxies at the same stellar mass, the set of galaxies undergoing RPS will have the same KS relation with the isolated control set at comparable gas surface densities. Individual galaxies may populate different $\Sigma_{\rm gas}$ ranges and hence occupy different subsets of the ensemble KS relation, which can be caused by both active (RPS-driven gas flows) and passive (gas consumption due to starvation) mechanisms, as shown in Figure \ref{fig:KS_same_slope_transition}. But there is no evidence of star formation efficiency change at given $\Sigma_{\rm gas}$ in the RPS cases. On the star formation main sequence plane ($\Sigma_{\rm SFR} - \Sigma_{*}$), the RPS galaxy disks (clear of tail contamination and inclination/projection effects) will show enhanced $\Sigma_{\rm SFR}$ above a certain $\Sigma_{*}$ threshold, and sparse sampling indicating disk truncation below the $\Sigma_{*}$ threshold. This is because galaxies undergoing RPS tend to have more centrally concentrated gas (and SFR) radial profiles than their isolated counterparts under starvation (e.g., Figure \ref{fig:radial_profile}). All predictions here assume that RPS is the only active effect and starvation is the only passive effect. We discuss how these assumptions are limited and their implications in \S \ref{subsec:limitations} below.

\subsection{Limitations}\label{subsec:limitations}
We made idealistic simplifications in our modeling choices in order to focus on the science goals. We adopted a single star formation and feedback recipe \citep{goldbaum_mass_2015,goldbaum_mass_2016} and a static dark matter potential, omitted the direct modeling of magnetic fields, turbulence, and cosmic rays, and only sampled a single (most probable; \citealt{wetzel_orbits_2011}) satellite orbit and a 45$^{\circ}$ wind inclination in each halo, instead of conducting a population study. In particular, we discuss the following two and their implications.

\textbf{(i) Gas removal by gravitational mechanisms.} Our controlled suite of hydrodynamical simulations only includes active gas removal by RPS; we are missing the gravitational mechanisms, including satellite-host and satellite-satellite interactions \citep{boselli_environmental_2006}. In clusters, RPS by the ICM is the dominant mechanism for cold gas stripping \citep{boselli_environmental_2006,cortese_dawes_2021}. In galaxy groups (lower relative velocities), satellite-satellite gravitational interactions are traditionally considered the primary stripping mechanism based on the observational evidence in various systems (e.g., \citealt{yun_high-resolution_1994,serra_discovery_2013,lee-waddell_wallaby_2019,wang_wallaby_2022}). However, the observational selection bias towards gas-rich galaxies in groups may have favored the gravitational mechanisms \citep{cortese_dawes_2021}; in fact, both simulations \citep{bekki_galactic_2014,bahe_star_2015,marasco_environmental_2016} and recent observational work \citep{roberts_lotss_2021,putman_gas_2021,kolcu_quantifying_2022} have found that RPS can be efficient in galaxy groups. RPS and gravitational (satellite-satellite) interactions are likely both effective in groups, and the relative importance depends on individual environments and satellites.

Missing the aspect of gravitational interactions, we likely overestimated the final $M_{\rm gas}$ (and $f_{\rm gas}$) in our galaxy group case (13W; Figures \ref{fig:gas_and_sfr_global} and \ref{fig:gas_fraction_time_evol}), as gravitational encounters can contribute to active gas removal. As the satellite still retains some ISM at 13W pericenter, gravitational interactions will additionally perturb the remaining gas, affecting its morphology and kinematics (Figures \ref{fig:mass_flow_via_density_slice} and \ref{fig:grav_potential_fallback_and_mixing}), and cause disturbances in the stellar disk. Such effects may also be present in the Milky Way and cluster halo cases but will have a weaker impact on the global properties (low likelihood of massive galaxy-galaxy close encounters in a Milky Way-like halo; high relative velocities in clusters).

Gravitational interactions will not change our key result of the RPS-induced star formation enhancement. Gas stripping by gravitational mechanisms is ``outside-in" like RPS and will have a relatively minimal impact on the dense gas in the disk center, where the SFR enhancement occurs in our simulations (e.g., Figures \ref{fig:radial_profile} and \ref{fig:local_sfr_star}). Our result is overall consistent with current observational evidence of triggered star formation that is of RPS origin, including global and local SFR enhancement \citep{vulcani_enhanced_2018,vulcani_gasp_2020,roberts_ram_2020}, and star formation in the tail \citep{hester_ic_2010,ebeling_jellyfish_2014,poggianti_gasp_2019}.

\textbf{(ii) Passive gas depletion and accretion.} We modeled passive gas depletion (i.e., free of direct removal) in our isolated galaxy simulation: steady consumption by star formation and stellar and supernova feedback-driven outflows. This scenario, however, only applies to a special case of galaxies where gas accretion has been halted, which we referred to as starvation following literature conventions \citep{larson_evolution_1980}. Accretion flows can naturally be halted by RPS, but they also likely occur in many of the star-forming field galaxies today, whose SFR relies on the rejuvenating cold gas accretion, e.g., via cooling inflows from the circumgalactic medium (CGM; \citealt{tumlinson_circumgalactic_2017}).

Gas accretion is a fundamental aspect of galaxy evolution that remains poorly constrained \citep{fox_gas_2017}. With the typical inflow rates and redshift dependence being highly uncertain, direct modeling of accretion is challenging. But we can infer the impact of accretion by comparing the SFR-time evolution in our simulated iso case (starvation without accretion), with the observational SFR-redshift relation \citep{speagle_highly_2014} at our modeled stellar mass. The difference in the SFR time evolution will characterize the star formation fueled by accretion missing in our iso simulation.

Taking the initial and final conditions in iso (Figure \ref{fig:gas_and_sfr_global}; $t_{\rm init}=500$ Myr and $t_{\rm final}=2980$ Myr), which converts to $t_{\rm lookback} \approx 2.5$ Gyr and the present day, the best-fit relation from \cite{speagle_highly_2014} gives $\log \rm SFR_{best-fit}$ of 0.07 and -0.24 dex. While in the iso simulation (starvation without accretion), the initial and final $\log \rm SFR_{iso}$ are 0.18 and -0.46 dex, respectively. The simulation values are still within the expected scatter of $\log \rm SFR_{best-fit}$ ($\pm$0.3 dex;  \citealt{speagle_highly_2014}), but unsurprisingly decrease faster with time due to the lack of replenishment. If we use the observational best-fit value as our field galaxy control case instead, which represents the average of observed star-forming galaxies at this stellar mass, the galaxy group case (13W) still shows globally enhanced SFR, although the enhancement becomes mild (a factor of 1.5 instead of $\sim$2.5).

\section{Summary and Conclusions}\label{sec:conclusion}

In this paper, we present a suite of galaxy-scale ``wind-tunnel" simulations with radiative cooling, star formation, and supernovae feedback, modeling a low-mass satellite galaxy undergoing RPS in various halo environments. The input time-varying ram pressure covers over three orders of magnitude (Figure \ref{fig:ram_pressure_evolution}), representing realistic satellite infall orbits from a Milky Way-like halo's $R_{200}$ to a cluster's pericenter. We simulate the same satellite galaxy in isolation for $\sim$3 Gyr as a control case, and compare the simulations in terms of their global evolution, gas morphology and kinematics, and the spatially resolved SFR-mass relations. Our key findings can be summarized as follows.

\begin{enumerate}[nolistsep]
    \item RPS has the potential to quench or enhance (up to a factor of $\sim$2.5) the global SFR of a satellite galaxy while gas is being removed (Figure \ref{fig:gas_and_sfr_global}). The impact on SFR depends on both the strength and the time derivative of the ram pressure.
    
    \item Star formation is radially centralized under a moderate (13W; group halo) or strong (14W; cluster halo) ram pressure profile, and it occurs in the stripped tail when the satellite reaches the cluster pericenter (Figure \ref{fig:sf_radius_and_height}). This is also reflected in a central enhancement of gas density (Figure \ref{fig:radial_profile}).
    
    \item Under an inclined wind (with face-on \textit{and} edge-on components), stripping in the disk outskirts dominates the gas mass loss, but when the pericentric ram pressure is insufficient for complete gas removal (13W), some stripped gas falls back and replenishes the central disk (Figures \ref{fig:grav_potential_fallback_and_mixing} and \ref{fig:mass_flow_mdot_series}). The edge-on component of ram pressure also drives a direct radial gas inflow (13W and 14W; Figure \ref{fig:cumulative_central_flows}) where it counters the disk rotation (right panel of Figure \ref{fig:grav_potential_fallback_and_mixing}).

\end{enumerate}
This radial gas transport has the following consequences:
\begin{enumerate}[nolistsep]
\setcounter{enumi}{3}
    
    \item The stripping set (13W and 14W) shows an excess of high $\Sigma_{\rm SFR}$-high $\Sigma_{\rm gas}$ material relative to the iso control set on the spatially resolved KS plane (Figure \ref{fig:local_KS_law}). However, the underlying KS relation is the same between the two sets when compared at similar $\Sigma_{\rm gas}$ (Figure \ref{fig:KS_same_slope_transition}), indicating that RPS has no direct effect on the star formation efficiency.
    
    \item On the spatially resolved SFR-stellar mass plane, the stripping set shows enhanced $\Sigma_{\rm SFR}$ at high $\Sigma_{*}$ (corresponding to central disk regions) relative to iso, and is truncated at the lowest $\Sigma_{*}$ (Figure \ref{fig:local_sfr_star}).

    \item The dense gas ratio ($R_{\Sigma_{10}}$; an approximation for $M_{\rm H_{2}}/M_{\rm H\,\textsc{i}}$) increases with time in the stripping set because of a combination of low-density gas removal and dense gas replenishment, as opposed to the decreasing trend in iso due to starvation (Figure \ref{fig:gas_fraction_time_evol}).

\end{enumerate}

Several of our findings agree with observational results. First, the RPS-induced global SFR enhancement is mild, up to a factor of $\sim$2.5 relative to iso, or $\sim$1.5 relative to the observational SFR-redshift relation (\citealt{speagle_highly_2014}; see \S \ref{subsec:limitations}). This agrees with the typical enhancement factor of $<$2 in observational samples of RPS-triggered star formation \citep{iglesias-paramo_tracing_2004,vulcani_enhanced_2018,roberts_ram_2020}. 
Second, despite occupying different $\Sigma_{\rm gas}$ ranges, the stripping and isolated sets follow the same local KS relation, consistent with \cite{jimenez-donaire_vertico_2023} (the VERTICO survey); the stripping set's $\Sigma_{\rm SFR}$ is smoothly enhanced at high $\Sigma_{*}$ --- the inner disk, consistent with \cite{vulcani_gasp_2020} (the GASP survey). Third, the stripping set acquires an enhanced dense gas ratio, which agrees with the high $M_{\rm H_{2}}/M_{\rm H\,\textsc{i}}$ ratios found for three GASP jellyfish galaxies \citep{moretti_high_2020}. Finally, RPS by a galaxy group medium can be effective for low-mass spiral galaxies and potentially lead to enhanced global SFR \citep{roberts_lotss_2021,kolcu_quantifying_2022}.

The radial redistribution of gas in the galaxy is a key result of this work: gas is stripped from the outskirts and enhanced in the center. It is the direct cause of the SFR enhancement when $P_{\rm ram}$ is insufficient to remove the entire gas disk; when $P_{\rm ram}$ is sufficient for central gas removal, the galaxy is ultimately quenched of star formation. This mass transport scenario is consistent with the increased dense gas ratio in jellyfish galaxies \citep{moretti_high_2020}, the local KS relation agreement between environmentally selected samples \citep{jimenez-donaire_vertico_2023}, and the comparable star formation efficiency 
between the leading and trailing halves of the disk (Figure \ref{fig:time_series_LH_TH}). We find that this explanation is a better match to the simulations than compression by ram pressure (\S \ref{subsec:discussion_RP_SF}) 
or a more efficient \HI-H$_{2}$ conversion (\S \ref{subsubsec:gas_fraction_discuss}) --- as the main driver for potential SFR enhancement.

\acknowledgements
We thank Dan Foreman-Mackey, Mary Putman, David Schiminovich, Jacqueline van Gorkom, and Ann Zabludoff for helpful discussions. We thank the anonymous referee for useful suggestions that improved the paper. JZ thanks Matthew Abruzzo, Nina Akerman, and Hui Li for the conversations on the simulations. ST thanks the GASP collaboration for useful conversations. GLB acknowledges support from the NSF (AST-2108470, XSEDE), a NASA TCAN award, and the Simons Foundation through the Learning the Universe Simons Collaboration. The simulations used in this work were run and analysed
on facilities supported by the Scientific Computing Core at the
Flatiron Institute, a division of the Simons Foundation. We also acknowledge computing resources from Columbia University's Shared Research Computing Facility project, which is supported by NIH Research Facility Improvement Grant 1G20RR030893-01, and associated funds from the New York State Empire State Development, Division of Science Technology and Innovation (NYSTAR) Contract C090171, both awarded April 15, 2010. Analyses of this work have made use of NumPy \citep{harris_array_2020}, Astropy \citep{astropy_collaboration_astropy_2013,astropy_collaboration_astropy_2018,astropy_collaboration_astropy_2022}, yt \citep{turk_yt_2011}, and Ipython \citep{perez_ipython_2007}.

\bibliography{RPS}
\bibliographystyle{aasjournal}
\end{document}